\documentclass[journal,10pt]{IEEEtran}

\usepackage{amsthm}
\usepackage{amsfonts}
\usepackage{amssymb}
\usepackage{mathrsfs}
\usepackage{mathtools}
\usepackage{float}
\usepackage[pdftex]{graphicx}
\usepackage{amsmath}
\usepackage{color}
\usepackage{multirow,multicol}
\usepackage{graphicx}

\usepackage{times}
\usepackage{textcomp}

\usepackage{verbatim}
\usepackage[table]{xcolor}% http://ctan.org/pkg/xcolor
\usepackage{balance}
\usepackage{lipsum}
\usepackage[inline]{enumitem}
\usepackage{cuted}
\usepackage[caption=false,font=footnotesize]{subfig}
\usepackage{cite}
\usepackage{hyperref}

\usepackage[table]{xcolor}% http://ctan.org/pkg/xcolor
\definecolor{intnull}{RGB}{213,229,255}
\definecolor{inteins}{RGB}{128,179,255}
\definecolor{intzwei}{RGB}{42,127,255}
\definecolor{intdrei}{RGB}{0,85,212}
\definecolor{intvier}{RGB}{0,51,128}
\definecolor{intfunf}{RGB}{0,34,85}

\DeclareMathOperator*{\minimize}{minimize\;} % thin space, limits underneath in 
\DeclareMathOperator*{\subjectto}{subject\;to\hspace{3pt}} % thin space, limits
\makeatletter
\newcommand{\smallsym}[2]{#1{\mathpalette\make@small@sym{#2}}}
\newcommand{\make@small@sym}[2]{%
	\vcenter{\hbox{$\m@th\downgrade@style#1#2$}}%
}
\newcommand{\downgrade@style}[1]{%
	\ifx#1\displaystyle\scriptstyle\else
	\ifx#1\textstyle\scriptstyle\else
	\scriptscriptstyle
	\fi\fi
}
\makeatother

\begin{document}
	%	\vspace{-30pt}
	\title{
		{			
			Near-Field Signal Processing: \\ Unleashing the Power of Proximity
	} }
	
	\author{\IEEEauthorblockN{ 
			Ahmet M. Elbir$^{1,2}$, \textit{Senior Member, IEEE,}
			\"{O}zlem Tu\u{g}fe Demir$^{3}$, \textit{Member, IEEE,} 
			Kumar Vijay Mishra$^{4}$, \textit{Senior Member, IEEE,} Symeon Chatzinotas$^2$, \textit{Fellow, IEEE} and Martin Haardt$^{5}$, \textit{Fellow, IEEE}
			\\
			{\footnotesize	$^{1}$Istinye University, 34396 Istanbul, Turkey \\
				$^{2}$University of Luxembourg, Luxembourg \\
				$^{3}$ TOBB University of Economics and Technology, Turkey \\
				{$^{4}$}United States DEVCOM Army Research Laboratory, Adelphi, USA\\
				$^5$Ilmenau University of Technology, Ilmenau, Germany }\\
			E-mail: {\footnotesize ahmetmelbir@ieee.org,
				ozlemtugfedemir@etu.edu.tr,
				kvm@ieee.org, symeon.chatzinotas@uni.lu, martin.haardt@tu-ilmenau.de  }
		}
%		\vspace{-50pt}
	}

	\maketitle
	
	\IEEEpeerreviewmaketitle

	\begin{abstract}
		After nearly a century of specialized applications in optics, remote sensing, and acoustics, the near-field (NF) electromagnetic propagation zone is experiencing a resurgence in research interest. This renewed attention is fueled by the emergence of promising applications in various fields such as wireless communications, holography, medical imaging, and quantum-inspired systems. Signal processing within NF sensing and wireless communications environments entails addressing issues related to extended scatterers, range-dependent beampatterns, spherical wavefronts, mutual coupling effects, and the presence of both reactive and radiative fields. Recent investigations have focused on these aspects in the context of extremely large arrays and wide bandwidths, giving rise to novel challenges in channel estimation, beamforming, beam training, sensing, and localization. While NF optics has a longstanding history, advancements in NF phase retrieval techniques and their applications have lately garnered significant research attention. Similarly, utilizing NF localization with acoustic arrays represents a contemporary extension of established principles in NF acoustic array signal processing. This article aims to provide an overview of state-of-the-art signal processing techniques within the NF domain, offering a comprehensive perspective on recent advances in diverse applications.
	\end{abstract}

	\vspace{-20pt}
	\section{Introduction}
\begin{figure}
		\centering
		{ \includegraphics[draft=false,width=\columnwidth]{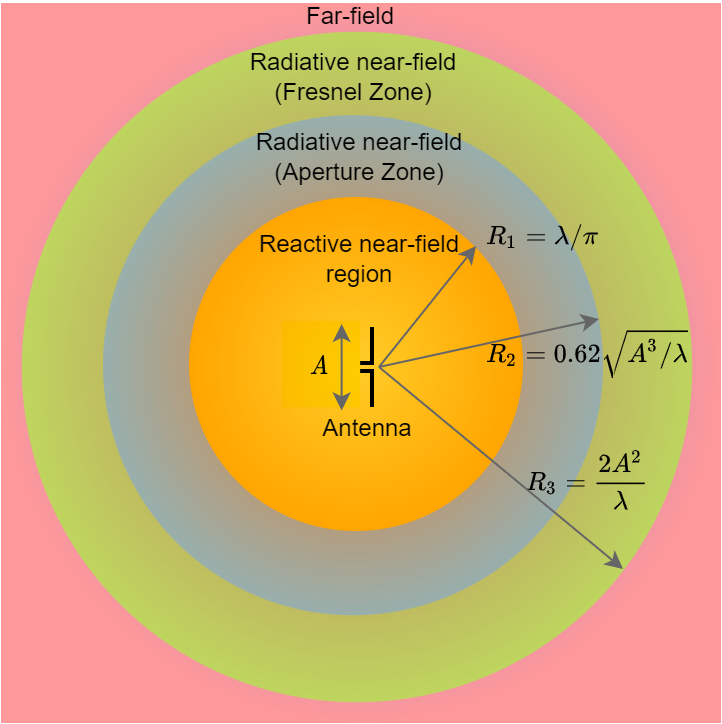} }

		\caption{EM field regions of an antenna of size $A$ for the signal wavelength $\lambda$. The reactive NF is immediately adjacent to the antenna and becomes negligible at  $R_1 =\lambda / \pi$. Beyond this boundary lies the radiative NF, itself split into two regions. The first part, known as the aperture zone, extends from $R_1$ to $R_2 = 0.62\sqrt{A^3/\lambda}$. The second part of the radiative NF is the Fresnel region, stretching from $R_2$ to $R_3=2 A^2 / \lambda$.  }
		%			\vspace*{-5mm}
		\label{fig_nf}
	\end{figure}

	Electromagnetic (EM) fields generated by an antenna create distinct spatial regions, each with unique characteristics \cite{balanis2011modern}. Closest to the antenna surface is the near-field (NF) region, which is divided into reactive and radiative sub-regions, as illustrated in Fig.~\ref{fig_nf}. {The reactive NF, immediately adjacent to the antenna, is dominated by fields that rapidly diminish with distance and become negligible beyond a boundary of approximately $\lambda / \pi$, where $\lambda$ is the operating wavelength~\cite{nf_book_Kalashnikov}. The intensity of the reactive fields is almost independent of the distance to the antenna, and the reactive fields become negligible in comparison with the radiative fields at a distance $\lambda/\pi$. Beyond this boundary lies the radiative NF, itself split into two regions. The first part, known as the aperture zone, extends from the reactive NF to approximately $0.62\sqrt{A^3/\lambda}$, where $A$ is the antenna's maximum linear size. In this zone, the reactive fields are  insignificant, the wavefront is quasi-planar, field strength remains nearly constant with distance, and the field amplitude distribution mirrors that of the antenna aperture. The second part of the radiative NF is the Fresnel region, stretching from $0.62\sqrt{A^3/\lambda}$ to $2 A^2 / \lambda$.} In this region, the radiation pattern begins to form, but its shape depends on the distance from the antenna. This is due to changing phase relationships and varying field amplitude ratios between different antenna elements as distance increases.

	%%-----------------------------------------------------
	%%-----------------------------------------------------

	As the observation point moves away from the antenna, the field amplitude initially oscillates before steadily decaying. At infinite distance, this attenuation becomes inversely proportional to the distance. Additionally, as distance increases, the phase and amplitude relationships between the fields from individual antenna elements gradually stabilize, and the angular distribution of the field becomes independent of distance. The outer boundary of the Fresnel zone, typically at $2 A^2 / \lambda$, marks the start of the far-field (FF) region, where the field distribution depends solely on the observation angle, and the field strength diminishes proportionally with distance. In this region, the phase  front of the EM waves  is spherical, though it appears planar within small angles. The internal parameters of the antenna relate to both reactive and radiative fields, while external parameters are concerned solely with the radiated EM waves.
	
	For much of its history, signal processing focused on FF applications, particularly in telecommunications, radar, and broadcasting, where the signals of interest were typically assumed to propagate over long distances with relatively simple wavefront characteristics. However, the unique NF properties were known for quite some time. Heinrich Hertz was among the first to observe that the field of a radiating dipole decays inversely with the third power of range at very close distances, rather than the inverse linear dependence observed in the FF. As a result, the field close to the dipole is significantly stronger than what one might expect from a simple extrapolation of the FF value. In 1909, Arnold Sommerfeld calculated the impact of NF perturbations on the radiation characteristics of a dipole antenna situated near the ground. 
	
	NF optics began to take shape in the late 1920s, driven by the need to surpass the diffraction limit in microscopy, enabling the visualization of structures at a nanometer scale \cite{pohl2012near}. This early work laid the groundwork for modern techniques such as NF scanning optical microscopy (NSOM), with more recent applications in %In optics, when the light passes through elements with subwavelength size features, such as diffractive elements, coded apertures, and diffusers, and the sensor is located near to these elements, NF optics models the propagation of light. Thus, NF optics finds applications in 
	%microscopy~\cite{optics_micros_Li2022Apr}, holography~\cite{optics_holo_Huhn2022Aug}, 
	Raman spectroscopic imaging~\cite{optics_raman_Kurouski2020Jun} and computational imaging~\cite{optics_computaImagin_Su2023Feb}. In the field of acoustics, the 1980s marked the beginning of a focused effort on NF applications, where sound waves interact with objects in close proximity, leading to advances in technologies like NF acoustic holography \cite{bai2013acoustic}, direction finding~\cite{acous_1_Wu2010Apr}, underwater 3-D localization~\cite{acoustics_localization_Shu2021Jun}, NF sonar \cite{lee2021direction}, beamforming~\cite{nf_acoustics_bf_Kumar2016Mar}, and sensor array optimization~\cite{acoustic_arrayOpt_Ryan2000Mar}. In ultrasound imaging, beamforming is usually performed via using wideband signals in NF~\cite{nf_US_beamforming_Viola2007Dec,nf_ultrasound2_He2015May}.
	
	In radar remote sensing, NF imaging has a rich research heritage dating back to the 1990s, particularly in applications like ground-penetrating radar, through-the-wall imaging, and synthetic aperture interferometry \cite{sarkar2008physics}. Recent NF radar signal processing applications focus on direction-of-arrival (DoA) estimation~\cite{nf_doaEst_MixedOrderStat_Molaei2021Dec},   localization, and correlated/coherent source localization~\cite{nf_coherent_Cheng2022Mar,elbir_farNear_nearFieldModel_Elbir2014Sep}. One of the major issues has been to derive the performance bounds, e.g., Cram\'{e}r-Rao bounds (CRB) for parameter estimation in NF scenarios~\cite{nf_crb_Wang2024Jan}. The consideration of a NF signal model makes the parameter estimation more challenging especially 
	when the received signal includes both NF and FF signals~\cite{nf_mixed_music_Liang2009Aug,nf_doaEst_mixed_Zheng2019Jul}. In order to estimate the DoA and the range information of the NF sources, a well-known technique is to employ a multiple signal classification (MUSIC) algorithm over both direction and range dimensions~\cite{music_Schmidt1986Mar}. As MUSIC is a subspace-based approach, it fails when the signals are coherent, i.e., fully correlated. In such cases, maximum-likelihood (ML)~\cite{nf_coherent_Cheng2022Mar} or compressed sensing (CS)~\cite{elbir_farNear_nearFieldModel_Elbir2014Sep} approaches can be employed. Relying on the high order statistics (HOS) of the array data, fourth order cumulants are also used~\cite{nf_HOS_Yuen1998Mar}. 
	
	More recently, the evolution of array signal processing in wireless communications is moving towards the use of small, densely packed sensors to create extremely large aperture arrays (ELAA), which significantly enhance angular resolution and beamforming gain~\cite{elbir2022Nov_SPM_beamforming}. Particularly, with the advent of sixth-generation (6G) networks, there is a notable adoption of ELAAs or surfaces, coupled with the exploitation of higher frequency bands like terahertz (THz) frequencies, shifting the EM diffraction field from the FF region to the NF. The extended array aperture and shorter wavelengths in the NF region, where the receiver is closer to the transmitter than the Fraunhofer distance~\cite{nf_primer_Bjornson}, result in non-planar signal wavefronts. %The NF region comprises reactive and radiative components. Given the array aperture $D$ and wavelength $\lambda$, these regions are delineated by the Fresnel distance ($\approx 0.62\sqrt{D^3/\lambda}$) and the phase variations within the Fresnel to Fraunhofer ($2D^2/\lambda$) distances. 
	With larger array apertures and frequencies, the NF range can extend tens to hundreds of meters, crucial for communication system design~\cite{survey_XL_MIMO_Wang2024Jan,nf_comm_survey3_Liu2023Aug,nf_comm_survey_short1_Cui2022Sep}. For instance, an antenna array of $1$-m aperture operating at $30$ GHz yields a NF region of up to $200$ meters, which is critical for system design in communication environments~\cite{nf_comm_survey_short1_Cui2022Sep}. In this context, spherical wavefront considerations become paramount in beamforming design, enabling signal focusing at specific 3-D locations rather than traditional FF beamsteering towards specific angles. The ELAA system facilitates not only directional signal reception but also transmitter localization, distinguishing it from conventional FF designs. However, exploiting the NF region presents both opportunities and challenges in communication~\cite{nf_comm_survey3_Liu2023Aug} and sensing~\cite{nf_crb_Wang2024Jan} applications. Beamforming in NF leverages depth information to enhance spatial multiplexing, catering to users at varying distances within the same direction. Additionally, high-resolution sensing and localization are imperative in the NF, necessitating innovative  communication strategies. Furthermore, in large  arrays operating at high bandwidths, the frequency-dependent array response poses challenges to conventional  beamforming, necessitating adaptive strategies for optimized beamforming gains~\cite{elbir_nf_radar_beamSquint_thz_Elbir2023Oct}.
	
	%	%%-----------------------------------------------------
	\begin{figure*}[t]
		\centering
		{\includegraphics[draft=false,width=.75\textwidth]{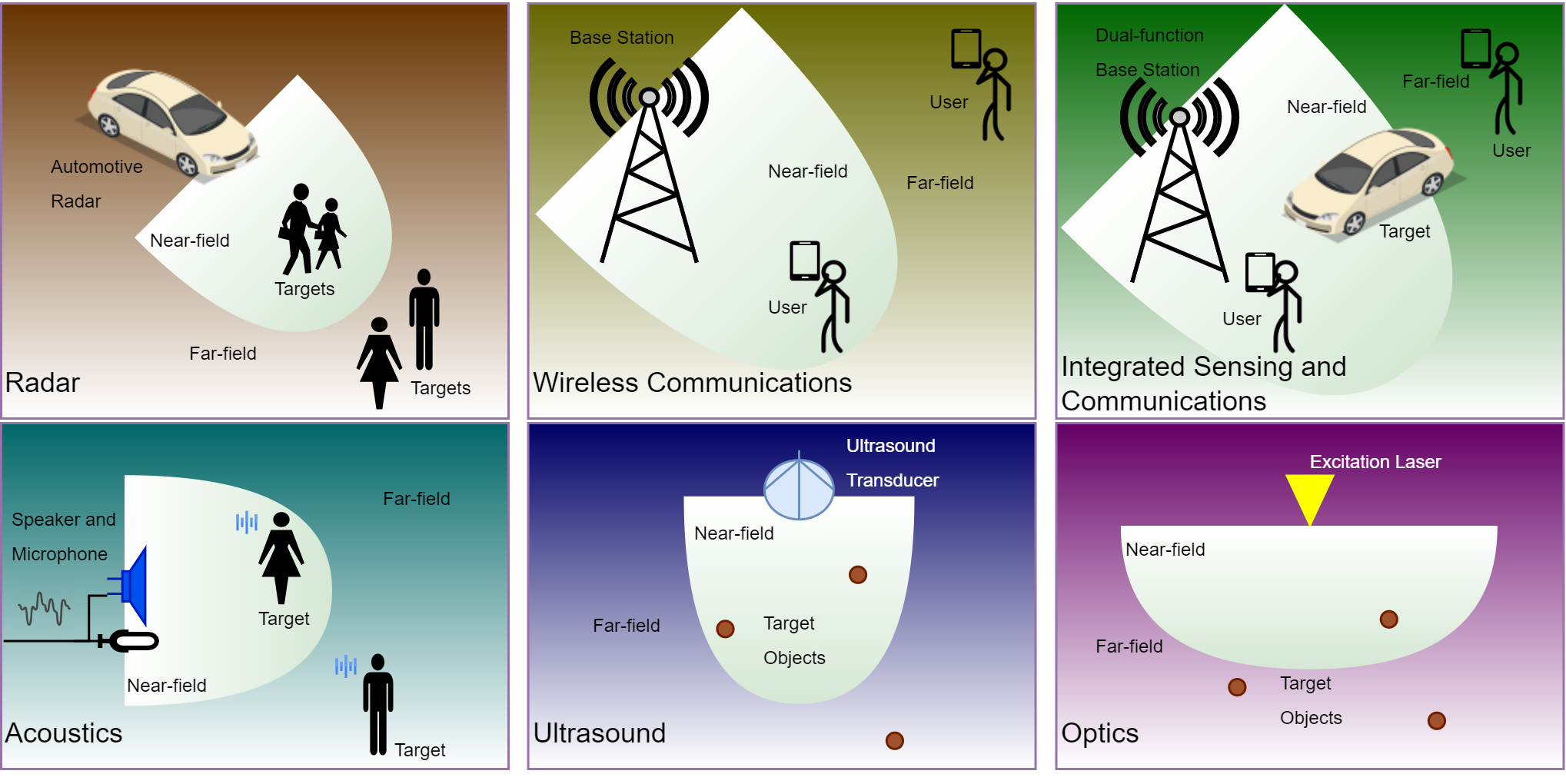} }
		\caption{Applications of NF signal processing on radar, communications, acoustics, ultrasound and optics.
		} 
		%			\vspace{-10pt}
		\label{fig_diag}
	\end{figure*}
	%	%%-----------------------------------------------------	
	
	%With the advances in multi-antenna communications as well as by increasing the carrier frequencies to mmWave and beyond, the NF phenomenon becomes an important area of research~\cite{survey_XL_MIMO_Wang2024Jan}. In particular, the user is more likely to fall in the NF of the base station (BS) as the frequency is higher and the array aperture is large sue to use of massive number of antennas at the BS to improve the beamforming gain. In order to accurately model the received signal in NF, there has been a substantial amount of research by both academics and industry. The major areas of research on NF wireless communications involve channel  estimation~\cite{nf_CE_3_LoSNLoS_Lu2023Mar} and beamforming~\cite{elbir_nf_ISAC_HB_Elbir2023Sep,nf_primer_Bjornson,nf_bf_inFOCUS_Myers2021Sep,nf_bf_OMP_Yang2023Mar,elbir_nba_omp_Elbir,nf_beamforming_Martin_Nwalozie}. The 
	
	Communications channel estimation in NF involves the reconstruction of the channel via estimating the complex channel gain as well as the direction and the range information of the dominant signal paths. In contrast to the radar scenario, the received signal includes several non-line-of-sight (NLoS) reflections, which may be in the NF of the receiver~\cite{nf_CE_3_LoSNLoS_Lu2023Mar,nf_bf_inFOCUS_Myers2021Sep}. This poses challenges to model the received signal for communications, wherein the aim is to improve the communication rate, rather than estimating the location of the user. In beamforming, the aim is to optimize the beamformer weights to maximize the communication rate. By taking into account the spherical-wave model of the received path signals, various techniques have been adopted for beamforming, e.g., CS~\cite{elbir_nf_ISAC_HB_Elbir2023Sep} and orthogonal matching pursuit (OMP). Combined with radar, integrated sensing and communications (ISAC) is a new paradigm to effectively use the system resources on a common platform for simultaneously performing radar sensing of targets or users as well as communication with the users~\cite{nf_isac_MUSIC__Wang2023May,elbir2021JointRadarComm}. In this respect, recent research also includes NF signal processing techniques to provide both sensing and communication functionalities, such as beamforming~\cite{elbir_nf_ISAC_HB_Elbir2023Sep} as well as user/target localization~\cite{nf_isac_MUSIC__Wang2023May}.

	The intent of this article is to encapsulate the \textit{recent} advancements in NF signal processing techniques. Notably, there is a distinct absence of surveys encapsulating signal processing strategies for the NF and acquainting with the nuances and challenges of this burgeoning field. {Near-field signal processing becomes relevant when the signal wavefront is spherical rather than planar, determined by the array aperture and wavelength. While the near-field concept remains fundamentally consistent across applications, the specific signal processing algorithms vary significantly depending on the task. For instance, just as near-field sensing and communications differ due to their unique requirements, similar variations exist between near-field acoustics, optics, and communications. 
		
		For example, in near-field acoustic source localization and beamforming, mode strengths depend on the source range, which is unknown \textit{a priori}. As a result, the conventional direct path dominance (DPD) test is conducted only in the time domain, unlike in the far-field case, where it spans both time and frequency. This constraint increases the number of frames required for near-field DPD when only time-domain data is available \cite{anderson2015spatial}. Furthermore, applying frequency smoothing in the near-field necessitates range-dependent normalization \cite{varanasi2019near}. Unlike far-field propagation, where source distance has little impact, near-field propagation exhibits significant energy variations at different observation points due to the changing distance from the source. Additionally, mode strengths are highly sensitive to noise and reverberation, which can degrade source localization performance and often require data-driven or learning-based methods for mitigation \cite{nadiri2014localization}.
		
		In near-field optics, common tasks include imaging and phase retrieval. Near-field imaging, which captures diffraction patterns using coherent light and a diffractive optical element, requires sensing matrices that combine Fourier and distance-dependent components, in contrast to the far-field scenario that uses only the Fourier matrix \cite{optics_phaseRetrival_KVM_Pinilla2023Feb}.  NF imaging in ultrasound is also common and used for specialized objects such as heterogeneous fluid excited by spherical waves. Fig.~\ref{fig_diag} illustrates NF signal processing across various domains. } 
	
	Lately, literature on NF signal processing involves short/extensive surveys~\cite{nf_comm_survey_short1_Cui2022Sep,nf_comm_survey3_Liu2023Aug,survey_XL_MIMO_Wang2024Jan} mainly on wireless communications while there is a substantial gap for an overview of NF signal processing techniques in radar, acoustics, ultrasound, and optics. {A comparison of the six use cases in Fig.~\ref{fig_diag} highlights the distinct challenges and future research directions in NF signal processing across various domains. Radar systems face challenges in precise localization and tracking of multiple targets in dynamic environments, motivating future research on robust algorithms and hardware for high-resolution sensing. Wireless communications must address beamforming and spatial multiplexing in NF regions, with opportunities to explore advanced array designs and near-field channel modeling. ISACs pose challenges in harmonizing radar and communication functionalities, driving research on joint waveform designs and resource allocation strategies. In acoustics, challenges include handling multi-path effects and environmental interference, suggesting future work on adaptive NF processing techniques for complex sound environments. Ultrasound applications need improved NF imaging accuracy and resolution, particularly for medical diagnostics, encouraging innovation in transducer design and signal reconstruction methods. Finally, optics involves challenges in NF sampling and wavefront manipulation for fine-detail imaging, paving the way for advancements in optical sensor technology and computational imaging algorithms. Overall, these comparisons underscore the need for domain-specific innovations while fostering interdisciplinary approaches to enhance NF signal processing across these diverse applications.} This article aims to present a synopsis of the state-of-the-art signal processing techniques in the NF with a focus on array processing and its contemporary applications. %In the remainder of this article, we first introduce the signal model for NF. Then,  we highlight the major signal processing techniques for radar, wireless communications, acoustics, ultrasound as well as optics. 

	\section{NF Signal Model}
\begin{figure}
		\centering
		{ \includegraphics[draft=false,width=.45\textwidth]{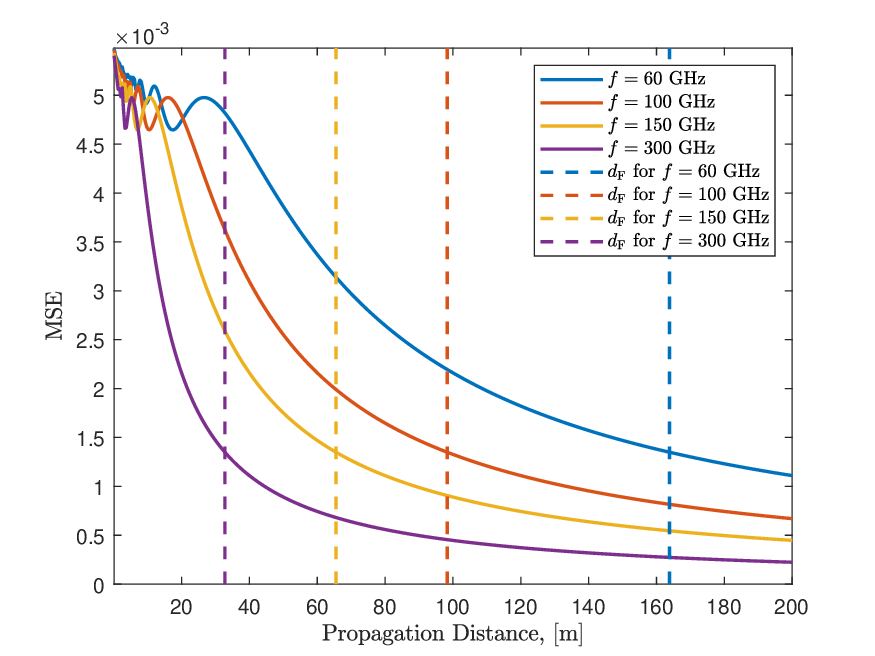} }
		\caption{MSE for the difference of NF and FF signal model for a ULA of $N=256$ elements on various frequencies.   }
		%			\vspace*{-5mm}
		\label{fig_MSE_distance}
	\end{figure}

	Consider a uniform linear array (ULA) with $N$ antenna elements and inter-element spacing $d$ that receives signals emitted from $K$ sources. We define $s_k(t)\in \mathbb{C}$ as the emitted signal from the $k$-th source with the direction of $\theta_k$ at time $t$. Then, the observation model at the $n$-th antenna element is given by
	\begin{align}
		y_n(t)= \sum_{k= 1}^{K} s_k(t)e^{-\mathrm{j}2\pi f_c \tau_{n,k}} + e_n(t),
		\label{sigModel1}
	\end{align}
	where $f_c$ is the carrier frequency and $e_n(t)\in \mathbb{C}$ denotes the additive white Gaussian noise, and {normalized} time delay $\tau_{n,k}$ is associated with  the $k$-th source signal propagation time among the antennas as
	\begin{align}
		\tau_{n,k} = \frac{r_k}{f_c \lambda} \left(\sqrt{ 1 + \frac{n^2 d^2}{r_k^2} - \frac{2 n d \sin\theta_k}{r_k}  }  -1 \right),
	\end{align}
	where $\lambda$ is the signal wavelength and $r_k$ denotes the range of the $k$-th source.
 {The aforementioned channel model primarily pertains to \textit{microwave} array signal processing in sensing and communications, whereas acoustics is sound wave and optics is light wave. While the optical and acoustic regimes have different models, they share key characteristics, including range dependence, spherical wave propagation, and distinct propagation zones.} {Depending on propagation distance $r$, the source location may be regarded in the reactive, radiative, or FF of the array~\cite[Ch.1]{nf_book_Kalashnikov}:}
	\begin{itemize}
		
			\item The reactive NF region: $0\leq r <\lambda/\pi$.      
			\item The first part of the radiative NF region (aperture zone): $\lambda/\pi \leq r <0.62\sqrt{D^3/\lambda}$, where $D$ denotes the aperture of the antenna array. Note that the aperture is the region occupied by the radiating antenna elements. Thus, its description applies to both antennas and array of antennas. In particular, the aperture of a ULA is defined as $D = (N-1)d$~\cite{balanis2011modern}. 
		
			\item The second part of the radiative NF region (Fresnel region): $ 0.62\sqrt{D^3/\lambda} < r < \frac{2D^2}{\lambda} $. It is in this region that the time delay $\tau_{n,k}$ can be approximated by using a Taylor series expansion as 
		
		\begin{align}
			\tau_{n,k} =-\frac{1}{2\pi f_c} \left(\omega_kn  + \kappa_k n^2+ O\left(\frac{d^2}{r_k^2}\right)\right),
		\end{align}
		where $\omega_k = \frac{2\pi d \sin \theta_k}{\lambda}$, $\kappa_k =- \frac{ \pi d^2 \cos
			^2 \theta_k}{\lambda r_k}$, {and $O(\nu)$ stands for the terms of order larger or equal to $\nu$}. Then, by neglecting $O\left(\frac{d^2}{r_k^2}\right)$ in the time delay expression, 
		the observation model in (\ref{sigModel1}) becomes
		\begin{align}
			y_n(t) = \sum_{k= 1}^K s_k(t) e^{\mathrm{j}(\omega_k n + \kappa_k n^2)  }  + e_n(t).
			\label{sig2}
		\end{align}
		
		\item The FF region starts after $d_\mathrm{F} = \frac{2D^2}{\lambda}$, which is called the Fraunhofer distance, and it covers $r> \frac{2D^2}{\lambda}$, wherein the plane wavefront can be approximated as planar with a maximum spherical-induced phase-shift of $\pi/8$ across the antennas~\cite{nf_primer_Bjornson}. Fig.~\ref{fig_MSE_distance} shows the mean-squared error (MSE) for the difference of NF and FF signal models for various frequencies and $d_\mathrm{F}$. We can see that this MSE decreases as the propagation distance $d$ increases and becomes negligible after $d > d_\mathrm{F}$. 
		
	\end{itemize}

	%%-----------------------------------------------------
	%%-----------------------------------------------------

	In  a compact form, the $N\times 1$ observation vector $\mathbf{y}(t)\in \mathbb{C}^N$  can be expressed as 
	\begin{align}
		\mathbf{y}(t) = \mathbf{A}\mathbf{s}(t) + \mathbf{e}(t),
	\end{align}
	where $\mathbf{y}(t) = \left[ y_1(t),\cdots, y_N(t)\right]^\textsf{T} $,  $\mathbf{e}(t) = \left[e_1(t),\cdots, e_N(t)\right]^\textsf{T}$ and $\mathbf{s}(t) = \left[s_1(t),\cdots, s_K(t)\right]^\textsf{T}$. Here, $\mathbf{A} = \left[ \mathbf{a}(r_1,\theta_1), \cdots, \mathbf{a}(r_K,\theta_K)\right]\in \mathbb{C}^{N\times K}$ denotes the array response, and $\mathbf{a}(r_k,\theta_k)\in \mathbb{C}^N$ represents the steering vector corresponding to the $k$-th source as
	\begin{align}
		\mathbf{a}(r_k,\theta_k) = \left[ e^{\mathrm{j}(\omega_k  + \kappa_k)  },\cdots, e^{\mathrm{j}(\omega_k N + \kappa_k N^2)  } \right]^\textsf{T}.
		\label{sv_radar1}
	\end{align}

	%%%%%%%%%%%%%%%%%%%%%%%%%%%%%%%%%%%%%%%%%%%%%%%%%%%%%%%%%%%%%%%%%%%%%%%%%%
	\section{NF Radar}
	{Extensive research has been conducted on the localization of radar signals in close proximity to antenna arrays.} This includes the estimation of the DoA angles as well as the range of the emitting source direction. Practical radar applications  involve several challenging scenarios, which requires advanced signal processing techniques for accurate parameters estimation.   %This section presents major signal processing techniques for NF radar scenarios including correlated signal estimation and performance bounds.
	Unlike FF scenario, the array response is range-dependent, which should be taken into account for source parameter estimation. Range-dependent beampattern is also observed in some far-field applications such as frequency diverse array (FDA) radars  and quantum Rydberg arrays \cite{vouras2023overview}. However, the wavefront is not spherical in these applications.
	
	\subsection{DoA Estimation and Localization}
	Subspace-based techniques, e.g., the MUSIC algorithm~\cite{music_Schmidt1986Mar}, have been widely used for NF DoA estimation and localization. Define $\mathbf{R}\in \mathbb{C}^{N\times N}$ as the sample covariance matrix of the array output $\mathbf{y}(t)$ as
	\begin{align}
		\mathbf{R} = \frac{1}{T} \sum_{t = 1}^{T} \mathbf{y}(t) \mathbf{y}^\textsf{H}(t),
	\end{align}
	where $T$ is the number of data snapshots collected from the array. Then, the MUSIC algorithm for NF DoA and range estimation is performed by finding the peak values of the following MUSIC spectra, i.e.,
	\begin{align}
		P(\theta,r) = \frac{1}{\mathbf{a}^\textsf{H}(\theta,r)\mathbf{U}_\mathrm{N}\mathbf{U}_\mathrm{N}^\textsf{H}\mathbf{a}(\theta,r)  },
	\end{align}
	where $\mathbf{U}_\mathrm{N}\in \mathbb{C}^{N\times(N-K)}$ denotes the noise subspace eigenvectors, which can be obtained from the eigenvalue decomposition of $\mathbf{R}$ as $\mathbf{R} = \mathbf{U}\boldsymbol{\Lambda}\mathbf{U}^\textsf{H}  $, where $\boldsymbol{\Lambda}\in \mathbb{C}^{{N\times N}}$ includes the eigenvalues of $\mathbf{R}$ in descending order and $\mathbf{U} = \left[\mathbf{U}_\mathrm{S}, \;\mathbf{U}_\mathrm{N}\right]\in \mathbb{C}^{N\times N}$  is composed of the eigenvectors corresponding to  the signal and noise subspaces, respectively.

	%%-----------------------------------------------------
	\begin{figure*}
		\centering
		{ \includegraphics[draft=false,width=.4\textwidth]{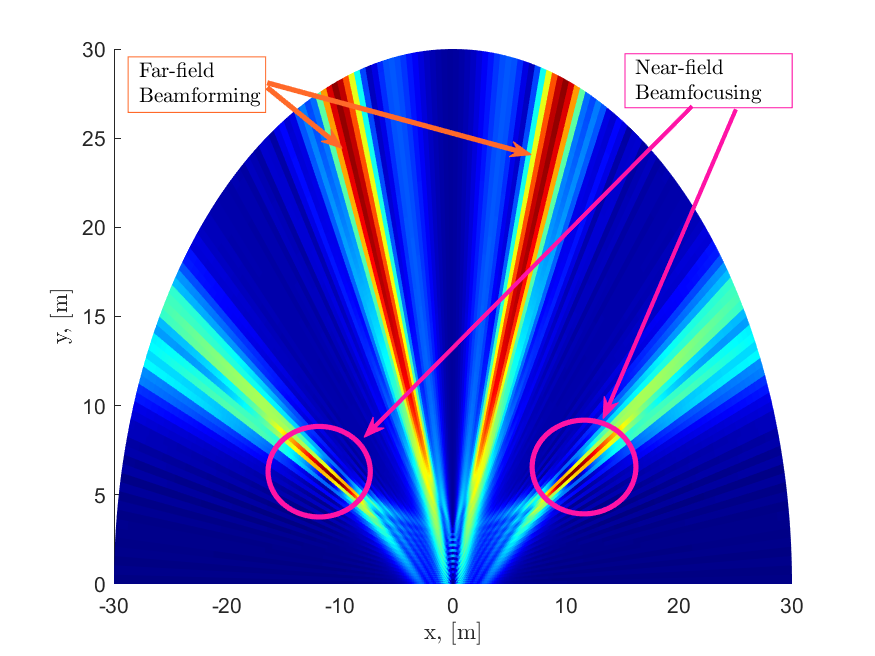} } 
		\caption{Array beampattern for the mixture of FF beamforming and NF beamfocusing.  }
		%			\vspace*{-5mm}
		\label{fig_coherent_mixed_far_near}
	\end{figure*}
	%%-----------------------------------------------------

	\subsection{Mixed NF and FF Source Localization}
	In some practical applications, the signal received by the array is a mixture of FF and NF sources~\cite{nf_doaEst_mixed_Zheng2019Jul}. Fig.~\ref{fig_coherent_mixed_far_near}(a) shows the array beampattern for mixture of FF and NF targets. For example, in high frequency (HF) radar, the NF multipaths have strong effects compared to very/ultra high frequency (V/U-HF) bands such that the received signal is composed of a FF source signal as well as its NF reflections~\cite{elbir_farNear_nearFieldModel_Elbir2014Sep}. Assume that there exist $K_\mathrm{F}$ FF and $K_\mathrm{N}$ NF emitting sources, for which the array output is given by
	\begin{align}
		\mathbf{y} (t) = \sum_{k = 1}^{K_\mathrm{F}} \mathbf{a}(\theta_k)s_k (t) + \sum_{k = K_\mathrm{F}+1}^{K}\mathbf{a}(\theta_k,r_k)s_k(t) + \mathbf{e}(t),
		\label{sigMixed}
	\end{align}
	where $K = K_\mathrm{F} + K_\mathrm{N}$ and $\mathbf{a}(\theta_k)\in \mathbb{C}^{N}$ denotes the FF steering vectors for $k \in \{1,\cdots, K_\mathrm{F}\}$. In this scenario, the MUSIC algorithm can be employed in a two-stage scheme to sequentially estimate FF and NF parameters. In~\cite{nf_mixed_music_Liang2009Aug}, two different special cumulant matrices are designed for this purpose. Consider a ULA with $N = 2\bar{N}+1$ elements and let $0$-th antenna be phase reference of the array. Then, the fourth-order cumulant of the array outputs can be expressed as 
	\begin{align}
		\mathrm{cum}\{y_p(t), y_{-p}^*(t), y_q(t),y_{-q}(t) \} = \sum_{k = 1}^{K} c_{s_k} e^{\mathrm{j}(2p\omega_k - 2q\omega_k  )  }  ,
	\end{align}
	where $p,q\in [-\bar{N},\bar{N}]$, and $c_{s_k} = \mathrm{cum}\{s_k(t), s_{k}^*(t), s_k^*(t),s_{k}(t) \}$ is the kurtosis of the $k$-th signal. Define $\bar{p} = p + \bar{N}+1$ and $\bar{q} = q + \bar{N}+1$. The $(\bar{p},\bar{q})$-th element of the cumulant matrix $\mathbf{C}_1\in \mathbb{C}^{N\times N}$ is 
	\begin{align}
		[\mathbf{C}_1]_{\bar{p},\bar{q}} =& \hspace{1pt}\mathrm{cum}\{y_{\bar{p}-\bar{N}-1  }(t), y_{-\bar{p}+\bar{N}+1}^*(t), y_{\bar{q}-\bar{N}-1}^*(t), y_{-\bar{q}+\bar{N}+1}(t)  \} \nonumber\\
		=& \sum_{k = 1}^K c_{s_k} e^{\mathrm{j}(2(\bar{p} - \bar{N} - 1)\omega_k - 2(\bar{q} -\bar{N}-1  )\omega_k ) }.
	\end{align}
	We can write $\mathbf{C}_1$ in a compact form as $\mathbf{C}_1 = \mathbf{B}_1\mathbf{C}_{s}\mathbf{B}_1^\textsf{H} $, where the virtual steering matrix and vectors are defined as $\mathbf{C}_{s} = \mathrm{diag}\{c_{s_1},\cdots, c_{s_K} \}\in \mathbb{C}^{K\times K}$, and $\mathbf{B}_1 = \left[\mathbf{b}_1(\omega_1),\cdots \mathbf{b}_1(\omega_K)\right]$, where $\mathbf{b}_1(\omega_k) = \left[e^{-\mathrm{j}2\bar{N}\omega_k },\cdots,e^{\mathrm{j}2\bar{N}\omega_k } \right]^\textsf{T}\in \mathbb{C}^{N}$ for $k = 1,\cdots, K$. By computing the MUSIC spectra based on the noise subspace eigenvectors of $\mathbf{C}_1$, the DoA information $\omega_k$ for $k \in \{1,\cdots, K_\mathrm{F}\}$ can be obtained via 
	\begin{align}
		P(\omega) = \frac{1}{\mathbf{b}_1^\textsf{H}(\omega){\mathbf{U}}_\mathrm{N_1}{\mathbf{U}}_\mathrm{N_1}^\textsf{H}\mathbf{b}_1(\omega)      },
	\end{align}
	where ${\mathbf{U}}_\mathrm{N_1}\in \mathbb{C}^{N\times (N-K)}$ denotes the noise subspace eigenvector matrix of $\mathbf{C}_1$. In order to estimate $\kappa_k$, another cumulant matrix $\mathbf{C}_2\in \mathbb{C}^{(4\bar{N}+1)\times (4\bar{N}+1)}$ is constructed as $\mathbf{C}_2 = \left[ \begin{array}{cc}
		\mathbf{C}_{2,1} & \mathbf{C}_{2,2} \\
		\mathbf{C}_{2,3} & \mathbf{C}_{2,4} 
	\end{array}\right]$, which can be written as $\mathbf{C}_2 = \mathbf{B}_2\mathbf{C}_{s}\mathbf{B}_{2}^\textsf{H}$, where  $\mathbf{B}_2 = \left[\mathbf{b}_2(\omega_1,\kappa_1),\cdots, \mathbf{b}_2(\omega_K,\kappa_K)\right]\in\mathbb{C}^{(4\bar{N}+1)\times K}$, and $\mathbf{b}_2(\omega_k,\kappa_k) = \mathbf{b}_2(\omega_k) \mathbf{b}_2(\kappa_k)\in \mathbb{C}^{4\bar{N}+1}$ (See~\cite{nf_mixed_music_Liang2009Aug} for the computation of $\mathbf{b}_2(\omega_k)$, $\mathbf{b}_2(\kappa_k)$ and $\mathbf{C}_{2,i}$, $i = 1,\cdots,4$). Then, by substituting the estimated $\omega_k$ into $\mathbf{b}_2(\omega_k,\kappa_k)$, the following MUSIC spectra reveals the range-dependent source parameters $\kappa_k$ for $k \in \{K_\mathrm{F} + 1,\cdots, K_\mathrm{F}+K_\mathrm{N}\}$, i.e.,
	\begin{align}
		P(\kappa) = \frac{1}{\mathbf{b}_2^\textsf{H}(\omega,\kappa){\mathbf{U}}_\mathrm{N_2}{\mathbf{U}}_\mathrm{N_2}^\textsf{H}\mathbf{b}_2(\omega,\kappa)      },
	\end{align}
	where ${\mathbf{U}}_\mathrm{N_2}\in \mathbb{C}^{(4\bar{N}+1)\times (4\bar{N}+1-K)}$ denotes the noise subspace eigenvector matrix of $\mathbf{C}_2$. 
	
	The aforementioned technique require multiple steps to sequentially estimate the DoA and range parameters of the mixture signals. Instead, more efficient techniques have been introduced in the literature by exploiting the special geometry/structure of the antenna arrays, e.g., subarrayed ULA and nested array~\cite{nf_doaEst_mixed_Zheng2019Jul}. 
	
	%Apart from these model-based approaches, data-driven techniques have also been devised~\cite{nf_mixed_DNN_Su2021Jul}.

	\subsection{Correlated/Coherent  Signal Estimation}
	In practice, the source signals are not always uncorrelated because of the received reflections from the objects which can be located in the FF or NF of the antenna array. In the most extreme scenario, the signals are coherent, i.e., fully correlated, such that the covariance matrix of the received signal turns out to be rank-deficient~\cite{nf_coherent_Cheng2022Mar}. For example, in multipath scenario,  the emitted signals from FF or NF sources are scattered from the objects in the vicinity of the antenna array. Consider a single FF source signal $s_1(t)$ (i.e., $K_\mathrm{F} = 1$), which is then reflected from $K_\mathrm{N} = K-1$ reflection points in the NF of the array. Thus, the reflected signals are coherent with $s_1(t)$, i.e, $s_k(t) = \zeta_k s_1(t) $  for $k \in \{2,\cdots, K\}$ and $\zeta_k \in \mathbb{C}$. Then, the mixture  model in (\ref{sigMixed}) is written for coherent signals as 
	\begin{align}
		\bar{\mathbf{y}} (t) = \mathbf{a}(\theta_1)s_1 (t) + \sum_{k = 2}^{K}  \mathbf{a}(\theta_k,r_k) {s_k(t)} + \mathbf{e}(t).
		\label{sigfar1Near}
	\end{align}
	Since the covariance matrix  of $\bar{\mathbf{y}}(t)$ is $\mathrm{rank}$-1, the subspace-based methods fail to resolve source angle and ranges. In~\cite{elbir_farNear_nearFieldModel_Elbir2014Sep}, a calibration-based approach is presented to estimate the angle and range parameters of mixture signals. Specifically, the NF signal components are first treated as disturbance signals to be calibrated, and the FF DoA angle is estimated. Then, the NF parameters are found. By exploiting the coherent signal model in (\ref{sigfar1Near}), {the $n$-th element of the observation $\bar{\mathbf{y}}(t)$  can be written as 
		\begin{align}
			\bar{y}_n (t) = {a}_n(\theta_1)s_1 (t) (1 + \beta_{n2} + \beta_{n3} + \cdots +  \beta_{nK}) + {e}_n(t),
		\end{align}
		where $\beta_{nk}\in \mathbb{C}$ is a direction-dependent parameter, and it is defined as 
		$\beta_{nk} = \frac{a_n(\theta_k,r_k)s_k(t)}{s_1(t)}$ for $k \in\{2,\dots, K\}$. Then, the array model for FF is 
		\begin{align}
			\mathbf{y} (t)= \boldsymbol{\Gamma} {\mathbf{a}(\theta_1)} s_1(t) + \mathbf{e}(t), 
	\end{align}}
	where $\boldsymbol{\Gamma} = \mathrm{diag} \{\gamma_1, \cdots, \gamma_N  \}\in \mathbb{C}^{N\times N}$ is a direction-dependent matrix representing the impact of NF signals as disturbances. As a result, the FF DoA angles can estimated with the knowledge of $\boldsymbol{\Gamma}$, which can be designed via a calibration technique. {Let $\tilde{\mathbf{Y}}(t) = \left[\tilde{\mathbf{y}}_1(t), \cdots,\tilde{\mathbf{y}}_C (t)  \right]\in \mathbb{C}^{N\times C}$ denotes the set of collected array measurements during the calibration for the calibration angles $ \{\tilde{\theta}_{1},\cdots, \tilde{\theta}_{C}  \}$. Suppose that the transmitted signal during the calibration $\tilde{s}(t)$ is known. Then, the FF steering vector at the direction $\tilde{\theta}_c$ is estimated as $\tilde{\mathbf{a}}(\tilde{\theta}_c) = \frac{\sum_{t =1}^{T}\tilde{y}(t) \tilde{s}^*(t)  }{\sum_{t=1}^{T}|\tilde{s}(t)|^2   }  $.         }
	Assume that the calibration data is collected uniformly in $[-\pi/2, \pi/2]$. The elements of the calibration matrix $\boldsymbol{\Gamma}_c$ at the $c$-th calibration angle $\tilde{\theta}_c$ can be found as $\gamma_{n,c} = \frac{ [\tilde{\mathbf{a}}(\tilde{\theta}_{c} )]_n  }{[\mathbf{a}(\theta_c)]_n   }$ for $c = 1,\cdots, C$ and $n = 1,\cdots, N$. Then, the FF DoA angle $\theta_1$ can be estimated from the following MUSIC spectra evaluated at $\tilde{\Theta}$ i.e., 
	\begin{align}
		P(\theta_c) = \frac{1}{\mathbf{a}^\textsf{H}(\theta_c)\tilde{\mathbf{U}}_\mathrm{N}\tilde{\mathbf{U}}_\mathrm{N}^\textsf{H}\mathbf{a}(\theta_c)      },
	\end{align}
	where $\tilde{\mathbf{U}}_\mathrm{N}\in \mathbb{C}^{N\times (N-1)}$ is the noise subspace eigenvector matrix for the covariance of $\mathbf{y}(t)$ in (\ref{sigfar1Near}). Once the FF signal are obtained, the array output in (\ref{sigfar1Near}) is transformed to the FF domain to perform angle estimation of the NF signals by employing a near-to-far transformation (NFT) matrix $\mathbf{T}\in \mathbb{C}^{N\times N}$, i.e.,  $\mathbf{A}_\mathrm{FF} = \mathbf{T}\mathbf{A}_\mathrm{NF}$, where $\mathbf{A}_\mathrm{FF}\in \mathbb{C}^{N\times C}$ and $\mathbf{A}_\mathrm{NF}\in \mathbb{C}^{N\times C}$ denote the collected FF and NF calibration measurements. Then, the transformed array output is given by
	\begin{align}
		{\mathbf{y}}_\mathrm{nf}(t) = \mathbf{T}\mathbf{y}(t) = \underbrace{\mathbf{T}\mathbf{a}(\theta_1)s(t) + \mathbf{T} \mathbf{e}(t)}_{\mathrm{distortions}} + \underbrace{\sum_{k = 2}^{K}\mathbf{T}\mathbf{a}(\theta_k,r_k) s_1(t) }_{\mathrm{desired \hspace{1pt}NF  \hspace{1pt}signal}},
		\label{sigNFT}
	\end{align}
	where the FF signal behaves like distortion. The transformed signal in (\ref{sigNFT}) is still rank-deficient. To alleviate this, forward-backward spatial smoothing (FBSS) technique~\cite{smoothing_Friedlander1992Apr} is used to construct a full-rank covariance matrix. Then, the DoA angles of the desired signals can be obtained via FF MUSIC algorithm based on the covariance of ${{\mathbf{y}}_\mathrm{nf}(t)}$. Finally, the estimation of the ranges of the NF signals can be performed via sparsity-based techniques by substituting the estimated FF and NF DoA angles~\cite{elbir_farNear_nearFieldModel_Elbir2014Sep}.

	%\cite{nf_coherent_Cheng2022Mar,elbir_farNear_nearFieldModel_Elbir2014Sep}

	%\subsection{Performance Bounds}
	
	%\cite{nf_crb_Khamidullina2021May,nf_crb_Wang2024Jan,nf_HOS_Yuen1998Mar}
	
	%%%%%%%%%%%%%%%%%%%%%%%%%%%%%%%%%%%%%%%%%%%%%%%%%%%%%%%%%%%%%%%%%%%%%%%%%%
	\section{NF Wireless Communications}

	NF signal processing %has been extensively studied in radar applications over the past three decades, it 
	is an emerging area of research in wireless communications, driven by advancements in mmWave~\cite{nf_comm_survey1_Lu2023Oct,nf_beamforming_Martin_Nwalozie} and THz~\cite{nf_CE_3_LoSNLoS_Lu2023Mar} technologies. For instance, mmWave massive {multiple-input multiple-output (MIMO)} systems, which incorporate a high number of antennas (e.g., $>128$) and operate at high frequencies (e.g., $>30$ GHz), can have a Fraunhofer distance extending up to hundreds of meters when employing a standard $\lambda/2$-spaced ULA configuration. In such cases, NF signal processing becomes crucial for performing key communication tasks such as channel estimation, beamforming, and resource allocation. %In this section, we highlight the state-of-the-art signal processing techniques for NF communications.

	\subsection{Channel Estimation}
	Consider $K$ single-antenna users and $L$ effective scatterers between each user and the {base station (BS)}. When all the effective scatterers between the BS and a particular user are located within the radiative NF of the BS, the channel between the BS and user $k$ can be expressed as
	\begin{align}
		\mathbf{h}_k= \sum_{l=1}^L\alpha_{k,l}\mathbf{a}(\theta_{k,l},r_{k,l}),
	\end{align}
	where $\alpha_{k,l}$ is the complex gain of the $l$-th path. The $\theta_{k,l}$ and $r_{k,l}$ are the corresponding angle and distance. To estimate the users' channels, the BS receives pilot sequences sent by the users during the training interval. To facilitate channel estimation and avoid interference, we assume that users are assigned orthogonal pilot sequences. This allows us to focus on a single user's received pilot signal (omitting the user index $k$), which is given by
	\begin{align}
		\mathbf{y} = \sqrt{\rho}\mathbf{A}\mathbf{h}+\mathbf{n},
	\end{align}
	where $\mathbf{y}$ is obtained after correlating the received signals with the pilot sequence of the user of interest, and $\rho$ is the pilot signal-to-noise ratio (SNR). The matrix $\mathbf{A}\in \mathbb{C}^{N_{\rm RF}\times N}$ represents the analog combining matrix, where the BS employs a hybrid beamforming architecture with $N_{\rm RF}$ RF chains. Each entry of $\mathbf{A}$ has a modulus of $1/\sqrt{N}$.  The independent additive noise is denoted by $\mathbf{n}\sim\mathcal{N}_{\mathbb{C}}(\mathbf{0},\mathbf{I}_N)$.
	
	To leverage the sparse characteristics of high-frequency channels when $L$ is relatively small, we can utilize OMP-type algorithms with codebooks specifically designed to account for the NF spherical wavefront \cite{nf_CE_3_LoSNLoS_Lu2023Mar}. For a good performance in the OMP algorithm, the dictionary matrix should exhibit low column coherence, meaning the magnitude of the maximum pairwise inner product between different columns is minimized. In FF scenarios, the dictionary matrix can be constructed using the columns of a discrete Fourier transform (DFT) matrix that correspond to FF array steering vectors. Conversely, for the NF case, a common approach to dictionary construction is the polar-domain design, which aims to achieve low column coherence, which is given as
	\begin{equation} \label{eq:column-coherence}
		\mu=\underset{p\neq q}{\max} \left|\mathbf{a}^\textsf{H}\left( \theta_{p}, r_{p}\right) \mathbf{a}\left( \theta_{q}, r_{q}\right)\right|  ,
	\end{equation}
	where $p$ and $q$ are the column indices of the dictionary. The absolute value of the inner product is 
	\begin{align}
		&\left|\mathbf{a}^\textsf{H}\left( \theta_{p}, r_{p}\right) \mathbf{a}\left( \theta_{q}, r_{q}\right)\right| \nonumber\\
		&= 	\left| \sum_{n=-\bar{N}}^{\bar{N}} e^{-\mathrm{j} \frac{2\pi}{\lambda}\left(dn(\sin\theta_q-\sin\theta_p)+d^2n^2\left(\frac{\cos^2 \theta_p }{2r_p} - \frac{\cos^2 \theta_q }{2r_q}\right) \right)} \right| ,   \label{eq:mag-inner}
	\end{align}
	where we have considered an ULA with $N = 2\bar{N}+1$ elements and $0$-th antenna being phase reference of the array. In polar-domain dictionary design, both angular and distance sampling are performed. For angular sampling, we consider two arbitrary locations where $\frac{\cos^2\theta}{r}=\frac{1}{\phi}$, with $\phi$ representing a constant corresponding to the distance ring $\phi$. On this distance ring, the nulls are obtained by sampling the angles
	\begin{align} \label{eq:angle-sampling}
		\theta = \arcsin\left( \frac{n\lambda }{N d}\right), \quad n=0,\pm 1, \pm 2, \ldots, \pm \left \lfloor \frac{Nd}{\lambda}\right \rfloor.
	\end{align}
	For distance sampling, we focus on two arbitrary locations with the same angle. Using the Fresnel integral approximation of the summation in \eqref{eq:mag-inner}, one can derive that  the distances should be sampled according to
	\begin{align}
		r = \frac{1}{s}\frac{N^2d^2\cos^2 \theta}{2\lambda \varepsilon^2}, \quad s=0,1,2,\ldots,
	\end{align}
	where $\varepsilon>0$ is a parameter that can be adjusted to guarantee a certain column coherence. For instance, to guarantee at most $0.5$ of the maximum inner product (i.e., $N$), $\varepsilon$ should be selected larger than 1.6. 
	
\begin{figure*}
		\centering
		\subfloat[]{ \includegraphics[draft=false,width=.45\textwidth]{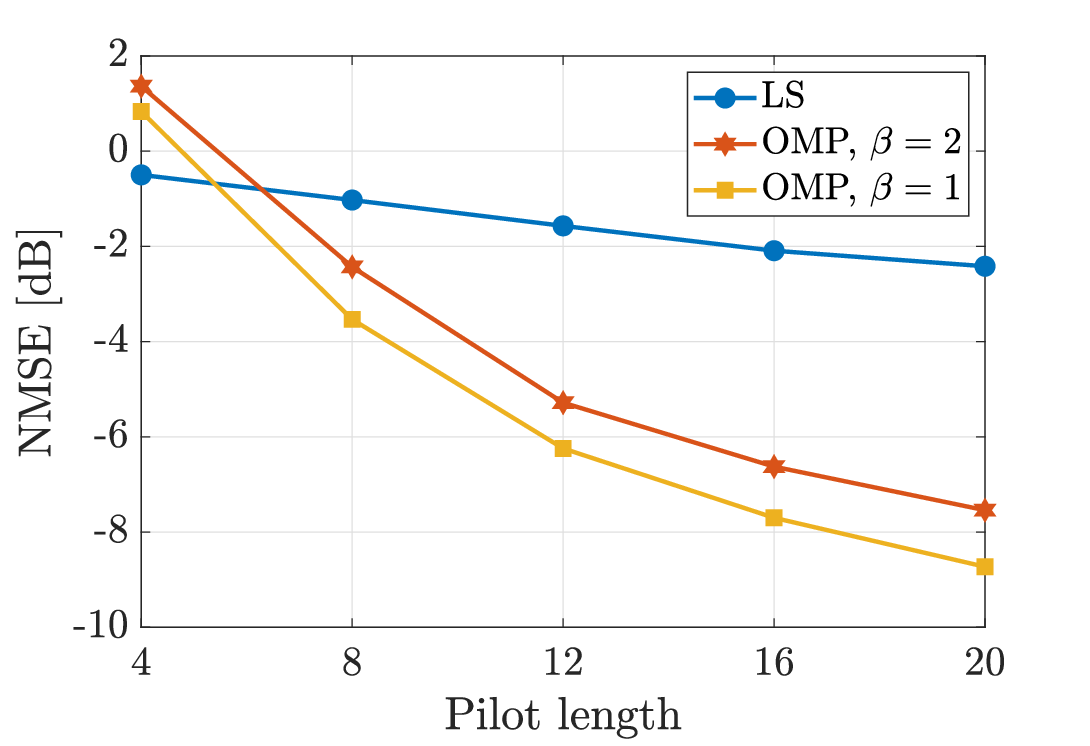} }
		\subfloat[]{ \includegraphics[draft=false,width=.45\textwidth]{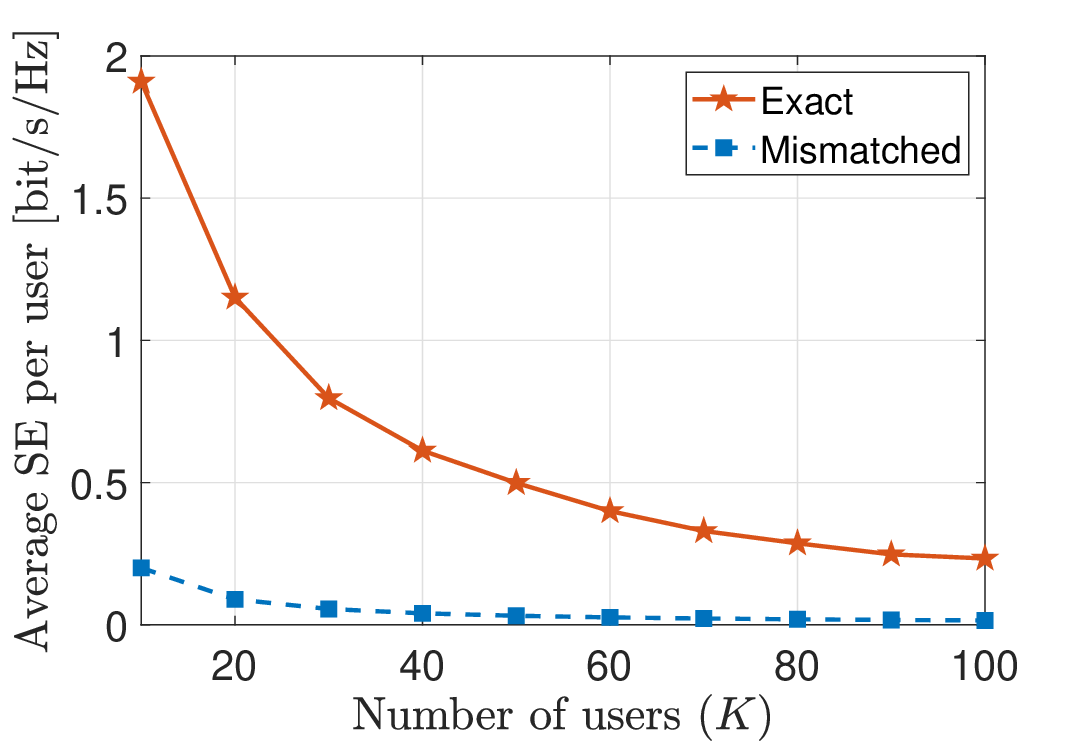} }  
		\caption{ Channel estimation and SE performance.  (a) NMSE versus pilot length for the LS and OMP-based channel estimators. (b) Average SE per user with exact NF channels and mismatched FF channels.   }
		%			\vspace*{-5mm}
		\label{fig_comm}
	\end{figure*}

	The dictionary matrix $\mathbf{W}$ is constructed using the array steering vectors $\mathbf{a}(\theta, r)$, with angles and distances selected according to the previously described angular and distance sampling methods to achieve the desired column coherence. In Fig.~\ref{fig_comm}(a), we compare the least squares (LS) channel estimator with the OMP algorithm using dictionaries designed with $\varepsilon = 1$ and $\varepsilon = 2$. The simulation parameters are as follows: $N = 256$, $N_{\rm RF} = 8$, and $\mathrm{SNR} = 10$ dB. As shown in the figure, as the pilot length increases, the normalized mean-squared error (NMSE) of all estimators decreases, with the OMP algorithm demonstrating superior performance when the pilot length is sufficiently large. This is because the OMP algorithm, combined with the polar-domain dictionary, effectively exploits both sparsity and the NF array steering vector structure, unlike the non-parametric LS estimator.
	
	NF effects can also be observed in midband frequencies for extremely large-scale arrays \cite{nf_primer_Bjornson}. At these frequencies, the sparsity may no longer exist. In such cases, the spatial correlation matrix can be utilized. When the full knowledge of the spatial correlation matrix is available, the optimal approach is to use the minimum mean-squared error (MMSE) estimator. However, acquiring the spatial correlation matrix becomes increasingly challenging, particularly as the array size grows. To improve upon the simplest LS alternative, one option for two-dimensional arrays is to leverage the low-rank characteristics of any plausible spatial correlation matrix, which is independent of specific user locations and only dependent on array geometry. The corresponding reduced-dimension LS (RS-LS) estimator can provide significantly better estimates than LS without requiring specific correlation matrix knowledge.

	%%-----------------------------------------------------
	%%-----------------------------------------------------

	\subsection{Beamforming vs. Beam-Focusing}
	
	Suppose a user has a line-of-sight (LoS) channel to the BS. If the user is situated in the FF of the BS and the BS employs maximum ratio transmission (MRT) precoding by adjusting the precoding vector according to the user's FF array steering vector, the resulting beampattern exhibits a finite beamwidth across the angular domain. However, along the distance at the user's angle, the beampattern remains strong. On the other hand, if the user is located in the NF of the BS, there is both a finite beam width across the angular domain and a finite beam depth along the distance domain. This can be understood by analyzing the NF beampattern at a specific angle $\theta$ by computing the normalized array gain:
	\begin{align}
		G(\theta,r)=   \frac{1}{N}\left | \mathbf{a}^\textsf{H}\left( \theta, r_0\right)\mathbf{a}\left( \theta, r\right)\right |,
	\end{align}
	where $r_0$ is the range of the user
	\cite{nf_comm_survey3_Liu2023Aug}. It can be shown that $G(\theta,r)$ can be approximated using Fresnel integrals $C(z)=\int_{0}^{z}\cos\left(\frac{\pi}{2}x^2\right)dx$ and $S(z)=\int_{0}^{z}\sin\left(\frac{\pi}{2}x^2\right)dx$ by
	\begin{align}
		G(\theta,r) \approx \left|\frac{C(z)+\mathrm{j}S(z)}{z}\right|  ,
	\end{align}
	where $z= \sqrt{\frac{N^2d^2\sin^2\theta}{2\lambda}\left|\frac{1}{r_0}-\frac{1}{r}\right|}$. Since $\left|\frac{C(z)+\mathrm{j}S(z)}{z}\right|$ is a decreasing function for $z\in[0,1.8]$ and it reduces to 0.5 when $z\approx 1.6$, there is a finite beam depth along the $r$ axis, which defines the 3 dB beam depth. After some mathematical manipulations, it can be shown that if $r<r_{\rm BD}$ where $r_{\rm BD}=\frac{N^2d^2\sin^2\theta}{2\lambda z_{\rm 3dB}^2}$, where $z_{\rm 3dB}=1.6$, the 3 dB beam depth (BD) is given as
	\begin{align}
		\mathrm{BD}_{\rm 3dB} = \frac{2r_0^2r_{\rm BD}}{r_{\rm BD}^2-r_0^2}.
	\end{align}
	On the other hand, if $r_0\geq r_{\rm BD}$, then the beam depth becomes infinity as in the FF case. 
	
	One of the significant implications of NF beam focusing is that multiple users can be simultaneously served by the BS, even if they are located along the same angle. This is not possible in the FF scenario, where nulling the interference between users is challenging. Due to the finite beam depth characteristics of beam-focusing, it is feasible to spatially multiplex many users, a concept known as \emph{massive spatial multiplexing} \cite{bjornson2024towards}.
	
	To exemplify massive spatial multiplexing, we consider the uplink of a multi-user MIMO system with $K$ single-antenna users. Following \cite[Sec.~II.B]{bjornson2024towards}, if one uses MMSE combining scheme, the spectral efficiency {(SE)} of the $k$-th user is given as
	\begin{align}
		\mathrm{SE}_k = \log_2\left(1+p_k\mathbf{h}_k^\textsf{H}\left(\sum_{i=1,i\neq k}^K p_i \mathbf{h}_i\mathbf{h}_i^\textsf{H}+\sigma^2\mathbf{I}_N \right)^{-1} \mathbf{h}_k\right),
	\end{align}
	where, $p_k$ denotes the transmit power of the $k$-th user, and $\sigma^2$ represents the noise variance. In Fig.~\ref{fig_comm}(b), we present the average SE per user for $N=512$, $f_c=30$ GHz, $p_k=0.2$ W, and $\sigma^2 = -87$ dBm, which corresponds to a 100 MHz bandwidth. $K$ users are positioned at an angle of $\theta=0$, with distances uniformly distributed between 20 and 500 meters. In the figure, the ``exact'' curve illustrates the SE derived using the precise NF LoS channels, while the ``mismatched'' curve represents the case where LoS channels are approximated by FF array steering vectors. As shown, approximating user channels as FF significantly reduces SE. This demonstrates that, despite identical path losses in both cases, utilizing NF channels leads to a considerable improvement in SE.

	\subsection{Wideband Processing and Beam-Squint}
	
\begin{figure*}
		\centering
		\subfloat[]{\includegraphics[draft=false,width=.30\textwidth]{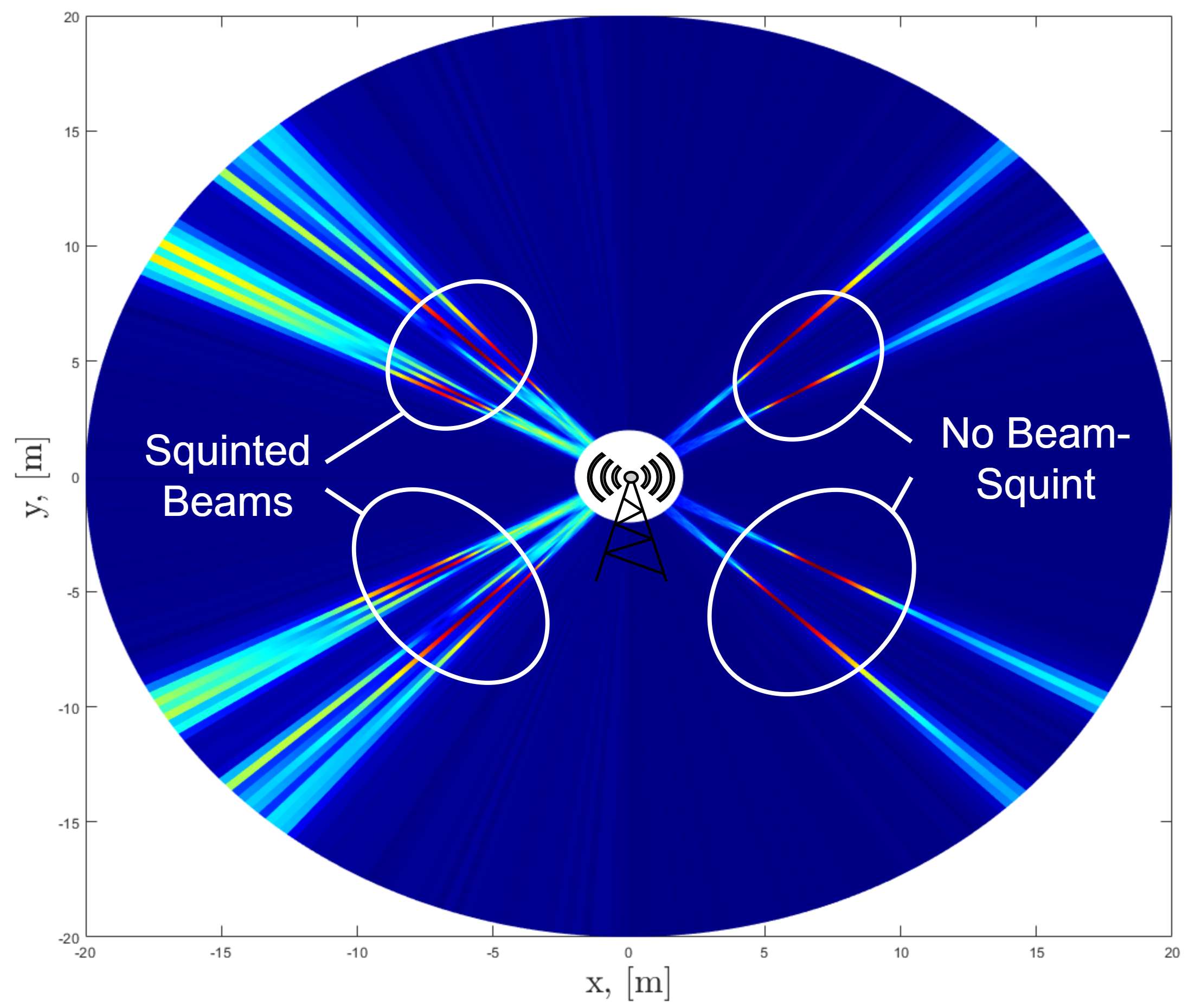} } 
		\subfloat[]{\includegraphics[draft=false,width=.35\textwidth]{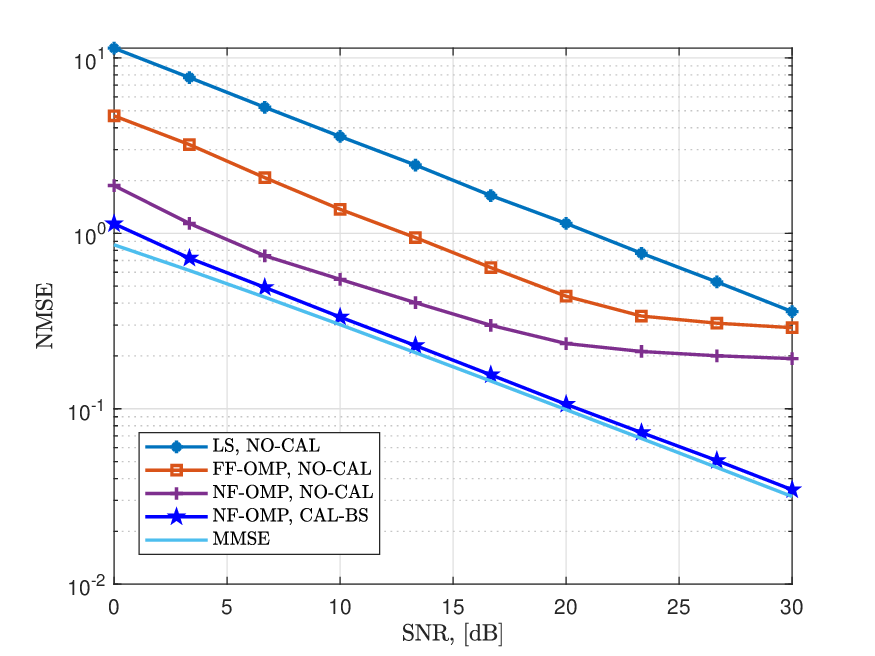} } 
		\subfloat[]{\includegraphics[draft=false,width=.35\textwidth]{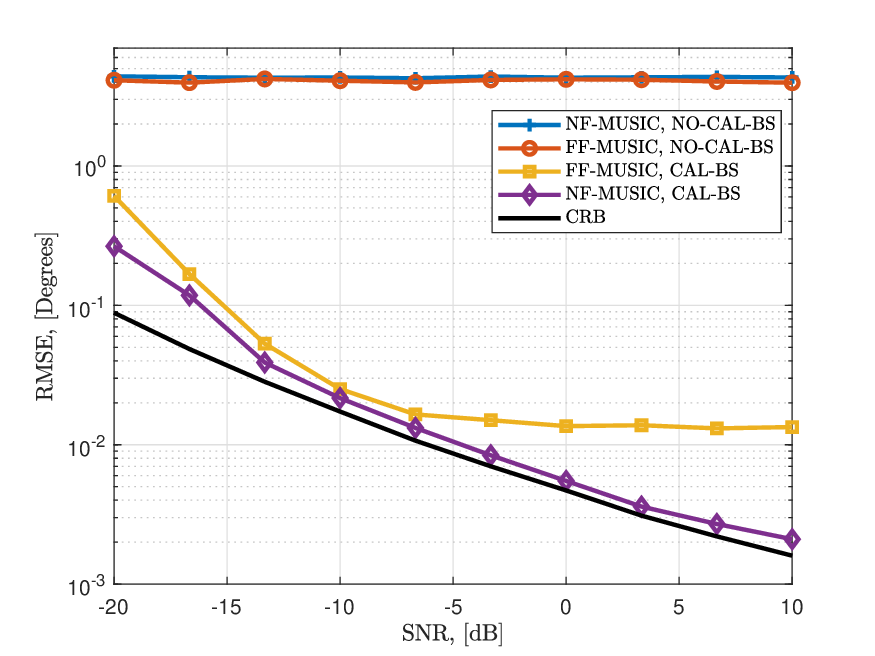} } 
		\caption{ Beam-squint calibration in NF: (a) {Array beampattern with and without beam-squint compensation}. (b) DoA estimation RMSE in NF radar. (b)  Channel estimation in NF communications.   	}
		%			\vspace*{-5mm}
		\label{fig_nf_beamSquint}
	\end{figure*}

	In wideband hybrid beamforming, the generated beam direction changes across the subcarriers compared to the central subcarrier. When the bandwidth is relatively large, e.g., in mmWave or THz designs, the beam direction squints especially at the high-end and low-end subcarriers. This phenomenon is called beam-squint~\cite{elbir_nf_radar_beamSquint_thz_Elbir2023Oct}. For instance, consider two users/targets located in the FF and NF at the broadside direction. In the FF, the angular deviation due to beam-squint is roughly $6^\circ$ ($0.4^\circ$) for $300$ GHz with $30$ GHz ($60$ GHz with $1$ GHz) bandwidth, respectively. However, NF beam-squint has approximately $(6^\circ, 10 \text{m} )$ angular and range deviation for a target/user located at $20$ m distance for $300$ GHz with $30$ GHz bandwidth. Fig.~\ref{fig_nf_beamSquint}(a) shows the array beampattern in the presence of NF beam-squint.

	%%-----------------------------------------------------
	%%-----------------------------------------------------

	Consider the array steering vector in (\ref{sv_radar1}), which can be rewritten in terms of the distance between the $k$-th source and the $n$-th antenna, $r_k^{(n)}$, as
	\begin{align}
		\label{steeringVec1}
		{\mathbf{a}(\vartheta_{k},r_{k}) =  [e^{- \mathrm{j}\frac{2\pi}{\lambda} r_{k}^{(1)} },\cdots,e^{- \mathrm{j}\frac{2\pi}{\lambda} r_{k}^{(N)} }]^\textsf{T},}
	\end{align}
	where  $\vartheta_k = \sin\theta_k$ and  $r_{k}^{(n)}= \sqrt{r_{k}^2  + (n-1)^2 d^2 - 2 r_{k}(n-1) d \vartheta_{k}   }$, which can be approximated~\cite{nf_primer_Bjornson,nf_comm_survey_short1_Cui2022Sep} as $	r_{k}^{(n)} \approx r_{k}  - (n-1) d \vartheta_{k}  + (n-1)^2 d^2 \Upsilon_{k}  ,$ where $\Upsilon_{k} = \frac{1- \vartheta_{k}^2}{2 r_{k}}$. Then, we can rewrite (\ref{steeringVec1}) as $	\mathbf{a}(\vartheta_{k},r_{k}) \approx e^{- \mathrm{j}2\pi \frac{f_c}{c_0}r_{k}} \tilde{\mathbf{a}}(\vartheta_{k},r_{k}),$ where the $n$-th element of $\tilde{\mathbf{a}}(\vartheta_{k},r_{k})\in \mathbb{C}^{N}$ is 
	\begin{align}
		\label{steeringVectorPhy2}
		[\tilde{\mathbf{a}}(\vartheta_{k},r_{k})]_n = e^{\mathrm{j} 2\pi \frac{f_c}{c_0}\left( (n-1)d\vartheta_{k}  - (n-1)^2 d^2 \Upsilon_{k}\right) }.
	\end{align}
	Due to beam-squint, the generated beam toward $({\vartheta}_{k},{r}_{k})$ deviates to the spatial location $(\bar{\vartheta}_{m,k},\bar{r}_{m,k})$ at the $m$-th subcarrier in the beamspace. Then, the $n$-th entry of the deviated steering vector in (\ref{steeringVectorPhy2}) for the spatial location is 
	\begin{align}
		\label{steeringVectorSpa}
		&[\tilde{\mathbf{a}}(\bar{\vartheta}_{m,k},\bar{r}_{m,k})]_n \hspace{-3pt}= \hspace{-2pt}e^{\mathrm{j} 2\pi \frac{f_m}{c_0}\left( (n-1)d\bar{\vartheta}_{m,k}  - (n-1)^2 d^2 \bar{\Upsilon}_{m,k}\right) },
	\end{align}
	for which we can finally define the NF beam-squint in terms of DoAs and ranges as
	\begin{align}
		\label{beamSplit2}
		\Delta(\vartheta_{k},m) &= \bar{\vartheta}_{m,k} - \vartheta_{k} = (\eta_m -1)\vartheta_{k}, \\
		\Delta(r_{k},m) &= \bar{r}_{m,k} - r_{k} = (\eta_m -1)r_{k}=  (\eta_m -1) \frac{1 - \eta_m^2 \vartheta_{k}^2}{\eta_m(1 -\vartheta_{k}^2)}r_{k},
	\end{align}
	where $\eta_m = \frac{f_c}{f_m}$ is the ratio of the central and $m$-th subcarrier frequencies.

	The observation of beam-squint differs when the receive array is in the NF, which is the area where the receive signal wavefront is spherical rather than plane-wave as in FF. In such a scenario, NF beam-squint causes the squint of the generated beam toward distinct locations rather than directions.  As a result, handling beam-squint in the NF is even more challenging than that in the FF since the impact of beam-squint and the model mismatch due to the spherical wavefront are intertwined. Thus, NF beam-squint brings up new research challenges in both radar and communications for target/user DoA estimation~\cite{elbir_nf_radar_beamSquint_thz_Elbir2023Oct},  beamforming~\cite{nf_primer_Bjornson,elbir2022Nov_SPM_beamforming}, waveform design, and resource allocation~\cite{nf_bf_inFOCUS_Myers2021Sep}. One direct solution, analogous to the FF methods, is to employ OMP or MUSIC~\cite{elbir_nf_radar_beamSquint_thz_Elbir2023Oct} algorithms with the dictionary of NF array responses for channel/DoA estimation and beamforming. Fig.\ref{fig_nf_beamSquint}(b-c) shows beam-squint-aware system performance for DoA estimation root mean-squared error (RMSE)  and channel estimation NMSE to account for both radar and communication scenarios, respectively~\cite{elbir_nf_radar_beamSquint_thz_Elbir2023Oct}. In both cases, it is clear that beam-squint should be accurately compensated as it leads to substantial performance loss in DoA/channel estimation. In comparison, the model mismatch, i.e., employing only FF model for the received signal, has relatively less impact on the performance.

	%\cite{nf_bf_wideband_Deshpande2022Dec,elbir_nba_omp_Elbir,elbir_nf_CE_Elbir2023Apr,elbir_nf_radar_beamSquint_thz_Elbir2023Oct,nf_bf_inFOCUS_Myers2021Sep}

	%\subsection{Performance Bounds}
	
	%\cite{nf_isac_bf_perfAnalysis_Zhao2024Jan,nf_perfAnalyiss_Chen2023Jun,nf_crb_Wang2024Jan,nf_perfAnalysis_Localization_Korso2010Feb}
	
	\subsection{Integrated Sensing \& Communications}
\begin{figure}
		\centering
		{\includegraphics[draft=false,width=\columnwidth]{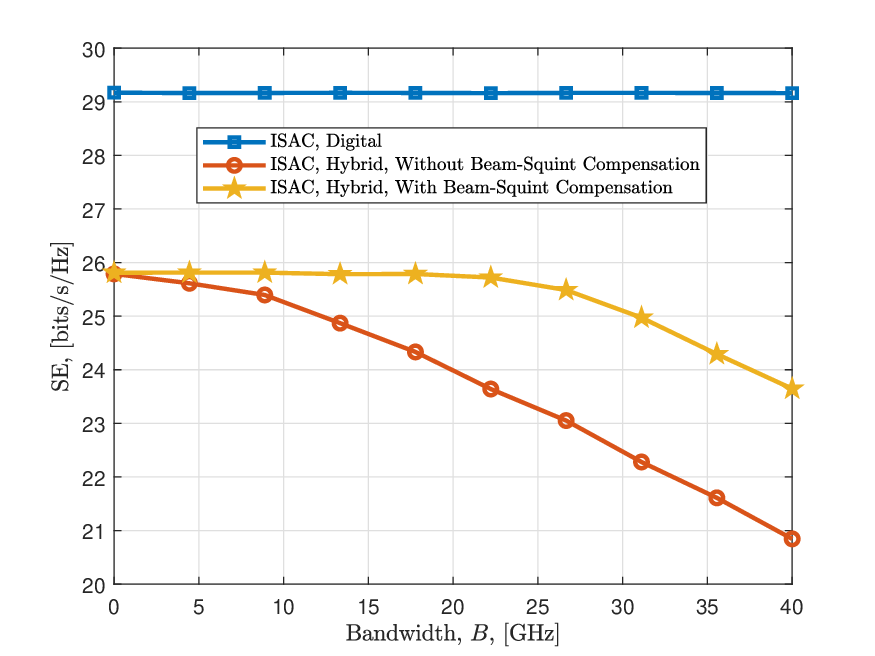} } 
		\caption{NF ISAC hybrid beamforming performance in terms of system bandwidth.  We assume $N=128$ transmit antennas with $N_\mathrm{RF}=8$ RF chains to generate beams toward $K=3$ targets and a user with $L=5$ NLoS paths. The  carrier frequency and the bandwidth are selected as $f_c = 300$ GHz and $B = 20$ GHz, respectively.  	}
		%			\vspace*{-5mm}
		\label{fig_ISAC_BF}
	\end{figure}

	Until recently, radar sensing and communication systems have been exclusively operated in non-overlapping frequency bands. However, a common demand for ubiquitous connectivity in wireless communications and high resolution radar sensing has led to a joint design of both systems in a shared spectrum as an effective solution: ISAC paradigms to share spectrum between radar and communications~\cite{elbir2021JointRadarComm}. As the combination of both scenarios, NF ISAC design involves simultaneously generating multiple beams toward both users and targets which are located in the NF~\cite{nf_isac_MUSIC__Wang2023May}. {Fig.~\ref{fig_ISAC_BF}  shows the NF ISAC beamforming performance in terms of the system bandwidth with and without beam-squint compensation. In digital beamforming, no beam-squint occurs as no analog components are used. Therefore, the digital beamforming performance is given as a benchmark. When hybrid analog/digital beamformers are used, the beam-squint should be compensated via either algorithmic or hardware-based techniques~\cite{elbir2021JointRadarComm,survey_XL_MIMO_Wang2024Jan}. We can see from Fig.~\ref{fig_ISAC_BF} that the beam-squint has a severe impact on the SE as the bandwidth becomes larger. Therefore, the NF beam-squint should be modeled and compensated accurately to maintain the satisfactory performance over a large bandwidth in the ISAC system.        }

	%%-----------------------------------------------------
	%%-----------------------------------------------------

	%	\cite{elbir_nf_ISAC_HB_Elbir2023Sep,nf_isac_bf_perfAnalysis_Zhao2024Jan,nf_isac_MUSIC__Wang2023May,elbir2022Aug_THz_ISAC}

	%%%%%%%%%%%%%%%%%%%%%%%%%%%%%%%%%%%%%%%%%%%%%%%%%%%%%%%%%%%%%%%%%%%%%%%%%%
	\section{Acoustics and Ultrasound}
	%Unlike the applications in radar and wireless communications, NF effect has been extensively studied 
	In acoustics and  ultrasound, the source location is usually in the NF of the sensor (microphone/hydrophone for acoustics and transducer for ultrasound) array, thereby observing the wavefront curvature. %We present most common NF applications in acoustics and ultrasound in the following.
	Fourier near-field acoustic holography (NAH) has been very popular since 1980s \cite{bai2013acoustic}. This technique reconstructs a 3-D sound field from the 2-D hologram scanned above the source surface in planar, cylindrical, and spherical geometries. This classical technique has been investigated for acoustic source localization for decades. Some recent extensions of NAH are discussed below.
	
	\subsection{Localization in Acoustics}
	Localization of  acoustic signals is a major task in air/underwater communications, security surveillance and sonar, wherein acoustic vector sensor (AVS) is employed as a measuring device~\cite{acoustics_localization_Shu2021Jun}. An AVS is composed of four components: three orthogonal velocity sensors measuring the Cartesian components and an isotropic pressure sensor measuring the acoustic pressure. Specifically, the $4\times 1$ NF array manifold for a AVS is given by
	\begin{align}
		\mathbf{v} = \left[\begin{array}{c}
			u \\
			v\\
			w\\
			p
		\end{array}\right] = \left[\begin{array}{c}
			\sin\varphi\cos\theta \\
			\sin \varphi \sin\theta \\
			\cos \varphi \\
			\frac{1}{\sqrt{1 + \left(\frac{\lambda}{2\pi r }\right)^2 }}e^{\mathrm{j}\arctan \frac{\lambda}{2\pi r}    }
		\end{array}\right],
		\label{acous1}
	\end{align}
	where $\theta$, $\varphi$ and $r$ represent azimuth/elevation angles and the distance of the source. The array manifold in (\ref{acous1}) can be estimated via the standard subspace-based methods such as estimation of signal parameters via rotational invariance (ESPRIT)~\cite{acous_1_Wu2010Apr}. Then, the 3D source location can be estimated by taking into account the NF propagation path attenuation and the phase difference among the sensors~\cite{acoustics_localization_Shu2021Jun}.

	%While NF signal processing has recently attracted great attention in wireless communications and localization, it has been extensively considered in other applications where the propagation distance is short, e.g., acoustics, ultrasound as well as optics. NF signal processing widely used in acoustics, wherein for example acoustics sensor arrays are employed for direction finding~\cite{acous_1_Wu2010Apr}, underwater 3-D localization~\cite{acoustics_localization_Shu2021Jun}, beamforming~\cite{nf_acoustics_bf_Kumar2016Mar,nf_bf_acoustics_He2012Jul} as well as sensor array optimization~\cite{acoustic_arrayOpt_Ryan2000Mar}. Also in ultrasound imaging, beamforming is performed via using wideband signals in NF~\cite{nf_US_beamforming_Viola2007Dec}.

	%			~\cite{acous_1_Wu2010Apr,acoustics_localization_Shu2021Jun}

	\subsection{Beamforming in Acoustics}
	In acoustics, beamforming is an important technology in microphone array signal processing to steer, shape and focus the acoustic waves toward a desired direction. A widely used beamforming technique is minimum variance distortionless response (MVDR), which is also called Capon beamforming~\cite{elbir2022Nov_SPM_beamforming}. Define $\mathbf{y}_\mathrm{m}(t)\in \mathbb{C}^N$ as the microphone array output, which are multiplied by the beamforming weights, i.e.,  $w_1,\cdots, w_N \in \mathbb{C}$. Then, the combined beamformer output is $			y_o(t) = \mathbf{w}^\textsf{H}\mathbf{y}_\mathrm{m}(t)$, where $\mathbf{y}_\mathrm{m}(t) = \mathbf{a}(\theta,r)s_\mathrm{m}(t) + \mathrm{e}_\mathrm{m}$ denotes the array output for the source signal $s_\mathrm{m}(t)\in \mathbb{C}$, and $\mathbf{w} = \left[w_1,\cdots, w_N\right]^\textsf{T}$ is the beamforming vector. For a NF acoustic source signal, the MVDR beamformer design problem is 
	\begin{align}
		\minimize_{\mathbf{w}}   \mathbf{w}^\textsf{H}\mathbf{R}_\mathrm{m}\mathbf{w} \hspace{5pt} \subjectto  \mathbf{w}^\textsf{H}\mathbf{a}(\theta,r) = 1,
		\label{ac_bf}
	\end{align}
	where $\mathbf{R}_\mathrm{m} = \frac{1}{T}\sum_{t = 1}^{T}\mathbf{y}_\mathrm{m}(t)\mathbf{y}_\mathrm{m}^\textsf{H}(t)$ is the sample covariance matrix of the array output and $\mathbf{a}(\theta,r)$ denotes the desired direction of interest of the beamformer. The optimal solution for (\ref{ac_bf}) is 
	\begin{align}
		\mathbf{w}_\mathrm{opt} = \left(\mathbf{a}^\textsf{H}(\theta,r)\mathbf{R}_\mathrm{m}^{-1}\mathbf{a}(\theta,r)\right)^{-1}\mathbf{R}_\mathrm{m}^{-1}\mathbf{a}(\theta,r),
	\end{align}
	which requires the knowledge of $\mathbf{a}(\theta,r)$. To relax this requirement, various beamforming techniques have been introduced for a more robust design~\cite{elbir2022Nov_SPM_beamforming}. Due to simplicity of the FF model, the transformation or calibration of the NF array output to FF have been widely adopted. Therefore, in order to deal with the NF effect in acoustics, a common approach is to employ a calibration technique by applying a gain/phase compensation to each array output so that the curved wavefront of the resulting signals appear as plane-wave~\cite{acoustic_arrayOpt_Ryan2000Mar}.	It is also shown in~\cite{acoustic_arrayOpt_Ryan2000Mar} that the improved beamforming performance can be achieved by exploiting the a priori knowledge of the distance between the source and the array, especially when the source of interest lies in NF while the interfering sources are located in the FF. Besides, subspace-based techniques, e.g., MUSIC and ESPRIT, can be employed for joint angle and range estimation~\cite{nf_acoustics_bf_Kumar2016Mar}.

	\subsection{Beamforming in Ultrasound Imaging}
	While beamforming has been widely used for narrowband signals in the FF, medical ultrasound imaging involves wideband signals originating in NF~\cite{nf_US_beamforming_Viola2007Dec}. Unlike classical radar imaging that assumes targets in FF as point sources, ultrasound NF imaging encounters extended scatterers. Hence, DoA estimation algorithms designed for point sources are inapplicable here. Generally, spread source modeling approach may be employed but the presence of severe background noise precludes widespread usage of this method. Hence, MVDR beamforming is commonly used by casting the problem as spatial spectrum estimation. However, this method is very sensitive to the errors in the imaging system as it is SNR-dependent based on the data covariance matrix. In order to overcome these challenges, sidelobe pattern control techniques have been introduced based on array optimization and diagonal loading~\cite{nf_ultrasound2_He2015May}. Alternatively, a full-scale electromagnetic or acoustic model of the scenario, including sensor/receiver characteristics and target environment, may be employed for DoA estimation.
	
	Define $\mathbf{y}_\mathrm{u}(x_p,y_p,z_p)\in \mathbb{C}^N$ as the NF response vector at the field point $(x_p,y_p,z_p)$ in Cartesian coordinate systems. Suppose that the mainlobe of the array is steering to the focus point $(x_f,y_f,z_f)$. Then, the combined beamformer output of the ultrasound transducer array from the acoustic point source at the $(x_p,y_p,z_p)$ becomes $B(x_p,y_p,z_p) =  \mathbf{w}^\textsf{H}\mathbf{y}_\mathrm{u}(x_p,y_p,z_p) $. The aim is to achieve maximum array gain at the desired location $(x_f,y_f,z_f)$ while minimizing the sidelobes in the beampattern. Thus, the beamforming design problem for sidelobe control is given by
	\begin{align}
		&\minimize_{\mathbf{w}} \xi 
		\nonumber \\
		& \subjectto \mathbf{w}^\textsf{H}\mathbf{y}_\mathrm{u}(x_f,y_f,z_f) = 1, 
		\nonumber \\
		& \|\mathbf{w}\| \leq \xi |\mathbf{w}^\textsf{H}\mathbf{y}_\mathrm{u}(x_p,y_p,z_p) |\leq \delta, 
	\end{align}
	where $ (x_p,y_p,z_p) \in \Theta_\mathrm{u},$ and $\delta$ controls the sidelobe level in the sidelobe region $\Theta_\mathrm{u}$. 
	%%%%%%%%%%%%%%%%%%%%%%%%%%%%%%%%%%%%%%%%%%%%%%%%%%%%%%%%%%%%%%%%%%%%%%%%%%
	\section{NF Optics}
	Until the 1980s, NF optics remained confined to microscopy. However, with the advent of applications such as tomography, cryogenic electron microscopy, and x-ray coherent diffractive imaging, NF techniques became highly diversified. In particular, infrared NF offers remarkable chemical sensitivity and nanoscopic spatial resolution, enabling the quantitative extraction of material properties from 3-D structured objects such as thin films and nanostructures \cite{govyadinov2014recovery}. A major advantage of NF imaging is its ability to surpass the resolution limits of conventional FF techniques, which are constrained by diffraction to about half the wavelength of the employed light. This limitation is a significant barrier to examining nanoscale objects, where resolutions in the range of $10$-$100$ nm are required. In contrast, NF propagation enables resolutions that are unattainable with traditional methods, especially in the holographic regime, where small Fresnel numbers produce high contrast. More recently, there has been a focus on NF phase retrieval (PR) for fields such as spanning holography, ptychography, and lens-less x-ray coherent diffractive imaging \cite{robisch2013phase}. 
	
	A general NF PR problem has the following setting. %In optical imaging applications, e.g.,  microscopy, holography, infrared imaging and Raman spectroscopy, the aim is to use light and the special properties of photons to obtain  detailed images of organs, tissues, and cells, by solving the wave equations to determine the field distribution of incident plane-wave over the object~\cite{opticsPhaseRetrival_Kornprobst2021Feb,optics_computaImagin_Su2023Feb}. Compared to the FF signal processing, the consideration of NF measurements provide enhanced nanoscopic spatial resolution and sensitivity, thereby exhibiting the improved quantitative extraction from the 3D objects~\cite{optics_phaseRetrival_KVM_Pinilla2023Feb}. 
	The spherical near-field samples are defined in the rotation space, which is a set of all possible rotations on the 3-D Euclidean space $\mathbb{R}^3$. This space is parameterized by three rotation angles: polarization angles, $\phi, \chi \in[0,2 \pi]$ and azimuth angle $\theta \in[0,\pi]$. Wigner D-functions are the orthonormal basis of this space, i.e., 
	\begin{align}
		\mathrm{D}_l^{k, n}(\theta, \phi, \chi)=N_l e^{-\mathrm{j} k \phi} \mathrm{d}_l^{k, n}(\cos \theta) e^{-\mathrm{j} n \chi},
	\end{align}
	where $\mathrm{d}_l^{k, n}(\cos \theta)$ is the Wigner D-function of band-limit degree $0 \leq l \leq B-1$ and orders $-l \leq k, n \leq l$, and $N_l=\sqrt{\frac{2 l+1}{8 \pi^2}}$ is the normalization factor. The spherical NF field $h(\theta, \phi, \chi)$ using a Wigner D-function expansion and bandwidth $B$ is
	\begin{align}\label{eq:expan}
		h(\theta, \phi, \chi)=\sum_{l=0}^{B-1} \sum_{k=-l}^l \sum_{n=-l}^l \alpha_l^{k, n} \mathrm{D}_l^{k, n}(\theta, \phi, \chi),
	\end{align}
	where $\left\{\alpha_l^n\right\}_{k, n=-l}^l$ are spherical mode coefficients to be reconstructed. Spherically sampling $\{\theta, \phi, \chi\}$ in $m$ points, the Wigner D expansion \eqref{eq:expan} is rewritten in the form of following linear feasibility problem: $\mathbf{h}=\mathbf{A}_W \boldsymbol{\alpha}$, where $\mathbf{h}=\left[h\left(\theta_1, \phi_1, \chi_1\right), \cdots, h\left(\theta_m, \phi_m, \chi_m\right)\right]^{\top}, \boldsymbol{\alpha} \in \mathbb{C}^n$ is constructed by Wigner D coefficients $\left\{\alpha_l^n\right\}_{k, n=-l}^l$, and the Wigner D-matrix $\mathbf{A}_W$ is 
	\begin{align}
		\mathbf{A}_W=\left(\begin{array}{cccc}
			\mathrm{D}_0^{0,0}\left(\theta_1, \phi_1, \chi_1\right) & \ldots & \mathrm{D}_{B-1}^{B-1, B-1}\left(\theta_1, \phi_1, \chi_1\right) & \\
			\vdots & \vdots & \ddots & \vdots \\
			\mathrm{D}_0^{0,0}\left(\theta_m, \phi_m, \chi_m\right) & \ldots & \mathrm{D}_{B-1}^{B-1, B-1}\left(\theta_m, \phi_m, \chi_m\right) &
		\end{array}\right).
	\end{align}
	This matrix is a collection of $m$ different samples of Wigner D-functions, where for each sample there exist Wigner D-functions related to its degree $l$ and order $|k|,|n|<B$. When only the phaseless measurements of NF radiation are available, then recovering $\boldsymbol{\alpha}\in \mathbb{C}^{n}$ from the embedded real-valued phaseless data $\boldsymbol{y}\in \mathbb{R}^{N}$ given the knowledge of the dictionary matrix $\mathbf{A}_W\in \mathbb{C}^{N\times Q}$ is the optimization problem 
	\begin{align}
		\text{ Find } \boldsymbol{\alpha} \hspace{1em} \text{ subject to } \hspace{1em}  \boldsymbol{y} = \lvert \mathbf{A}_{W}\boldsymbol{\alpha} \rvert,
	\end{align}
	where the non-convex constraint models the measurement process. This PR problem may also be generalized to mixed measurements from near-, middle-, and far-zone fields \cite{optics_phaseRetrival_KVM_Pinilla2023Feb}. Since the problem results in a non-convex inverse problem, the recovery method to obtain the phase generally exploit a specific optical design or signal structure \cite{optics_phaseRetrival_KVM_Pinilla2023Feb}.

				%%%%%%%%%%%%%%%%%%%%%%%%%%%%%%%%%%%%%%%%%%%%%%%%%%%%%%%%%%%%%%%%%%%%%%%%%%
				\section{Summary and Future Outlook}
				NF signal processing, traditionally regarded as a specialized niche, has a rich and extensive history that has significantly contributed to the advancement of various scientific fields, including electromagnetics, acoustics, and medical imaging. Classical NF techniques laid the groundwork for understanding and manipulating electromagnetic fields at close ranges, enabling breakthroughs that have shaped modern technology. In recent years, however, there has been a remarkable resurgence of interest in NF signal processing, driven by advancements in emerging technologies such as ELAA and ISAC. Further, enhanced capabilities for conducting measurements across various optical regimes have reinvigorated NF optics, leading to a proliferation of specialized applications. Recent quantum-inspired techniques like those utilizing Rydberg atoms have underscored the need to revisit and expand NF signal processing approaches. Nearly all near-field applications require addressing range dependence and spherical wavefronts. However, the specific signal processing tasks vary significantly across domains like wireless communications, optics, and acoustics. For instance, near-field beamforming in acoustics shares structural similarities in steering vectors with wireless communications. In acoustics, these steering vectors are often replaced with acoustic transfer functions to account for room acoustics, whereas near-field beamforming in communications relies on analytical models. Similarly, near-field optics primarily focuses on imaging or scene reconstruction, yet its propagation-zone dependence parallels that of near-field sensing. In summary, the ongoing exploration of NF theory and its applications holds great promise for addressing new and complex problems in signal processing.

				\balance
				%	\vspace{-1em}
				\bibliographystyle{IEEEtran}
				%	\bibliography{references_128}
				\bibliography{references_135}

						\begin{IEEEbiographynophoto} {Ahmet M. Elbir} (Senior Member, IEEE) received the Ph.D. degree in electrical engineering from Middle East Technical University in 2016. He is currently an Associate Professor at Istinye University and a Research Fellow at University of Luxembourg.
						\end{IEEEbiographynophoto}
						
							\begin{IEEEbiographynophoto} {\"{O}zlem Tu\u{g}fe Demir} (Member, IEEE) received the B.Sc., M.Sc., and Ph.D. degrees in electrical and electronics engineering from Middle East Technical University, Ankara, Turkey, in 2012, 2014, and 2018, respectively. She is currently an Assistant Professor at TOBB University of Economics and Technology.
						\end{IEEEbiographynophoto}
						
							\begin{IEEEbiographynophoto} {Kumar Vijay Mishra} (Senior Member, IEEE) obtained a Ph.D. in electrical engineering and M.S. in mathematics from The University of Iowa in 2015, M.S. in electrical engineering from Colorado State University in 2012, and B. Tech. summa cum laude (Gold Medal, Honors) in electronics and communication engineering from the National Institute of Technology, Hamirpur (NITH), India in 2003. He is currently Senior Fellow at the United States Army Research Laboratory.
						\end{IEEEbiographynophoto}
						
							\begin{IEEEbiographynophoto} {Symeon Chatzinotas} (Fellow, IEEE) received the M.Eng. degree in telecommunications from the Aristotle University of Thessaloniki, Thessaloniki, Greece, in 2003, and the M.Sc. and Ph.D. degrees in electronic engineering from the University of Surrey, Surrey, U.K., in 2006 and 2009, respectively. He is currently a Full Professor, the Chief Scientist I, and the Head of the SIGCOM Research Group, SnT, University of Luxembourg.
						\end{IEEEbiographynophoto}
						
							\begin{IEEEbiographynophoto} {Martin Haardt} (Fellow, IEEE) received his Diplom-Ingenieur (M.S.) degree from the Ruhr-University Bochum in 1991 and his Doktor-Ingenieur	(Ph.D.) degree from Munich University of Technology in 1996 after studying	electrical engineering at the Ruhr-University Bochum, Germany, and at Purdue
								University, USA. He is currently a Full Professor in the Department of Electrical Engineering and Information Technology and Head of the Communications Research Laboratory at Ilmenau University of Technology, Germany.
						\end{IEEEbiographynophoto}

			\end{document}